# Effect of Horizontal Spacing and Tilt Angle on Thermo-Buoyant Natural Convection from Two Horizontally Aligned Square Cylinders

**Subhasisa Rath*[1], Sukanta Kumar Dash[2]**

[1]School of Energy Science & Engineering, Indian Institute of Technology Kharagpur, 721302, India

[2]Department of Mechanical Engineering, Indian Institute of Technology Kharagpur, 721302, India

**\* Corresponding Author**
E-mail: subhasisa.rath@gmail.com
Tel.: +91-9437255862
ORCID ID: https://orcid.org/0000-0002-4202-7434

**ABSTRACT**

Laminar natural convection heat transfer from two horizontally aligned square cylinders has been investigated numerically using a finite-volume method (FVM) approach. Computations were performed to delineate the momentum and heat transfer characteristics under the following ranges of parameters: horizontal spacing between the cylinders ($0 \leq S/W \leq 10$), tilt angle of the square cylinder ($0^0 \leq \delta \leq 60^0$), and Grashof number ($10 \leq Gr \leq 10^5$) for some specific Newtonian fluids having Prandtl number ($0.71 \leq Pr \leq 7$). The comprehensive results are represented in terms of temperature contours and streamlines, velocity and temperature profiles, the mass flow rate in the passage between the cylinders, local and average $Nu$, and the drag coefficient. Owing to the development of a chimney effect, the heat transfer increases with decrease in the horizontal spacing up to a certain limit, whereas it significantly degrades with a further decrease in the spacing. The square cylinder having $\delta = 45^0$ shows a higher heat transfer, whereas it is least for $\delta = 0^0$. At higher $Gr$ and $Pr$, the average $Nu$ is found to be in excess of 22% at $\delta = 45^0$ compared to at $\delta = 0^0$. Overall, the average $Nu$ has a strong dependence on both $Gr$ and $Pr$, whereas it is a weak function of $S/W$ and $\delta$. Furthermore, the entropy generation is reproduced non-dimensionally in terms of the Bejan number. Finally, a correlation for the average $Nu$ has been developed, which can be useful for the engineering calculations.

## 1. Introduction

The thermo-buoyant natural convection heat transfer has emerging potential applications in the field of engineering as well as in real life practical scenarios. The temperature dependent density that drives the flow in natural convection need not require any external mechanical forcing agent, which makes the complete system devote of any complexity, reliable and cheap. However, from a theoretical standpoint, the coupled momentum and energy equations are to be solved simultaneously in computational fluid dynamics (CFD) to elucidate both momentum and heat transfer characteristics, which appends additional complexity in the solution. Owing to the pragmatic significance over the last fifty years or so, free convection from various shapes of bluff bodies in air has been investigated experimentally, analytically, and numerically in both confined and unconfined fluid medium [1-5]. However, due to high demand for the knowledge of transport phenomenon, natural convection plays a vital role in present days as well.

Natural convection from a square cylinder denotes numerous industrial applications including heat exchanger design, electronic cooling, heating or drying of plywood sheets thermal treatment of various food-stuffs, convective drying in fluidized bed, etc. Zeitoun and Ali [6] numerically investigated the 2D laminar natural convection heat transfer from a horizontal rectangular duct of varying aspect ratio including a square cylinder. Recirculation zones were seen to form at the top of the rectangular duct at higher aspect ratio. Mahmud et al. [7] numerically demonstrated the orientation effect of an isothermal square cylinder on laminar natural convection heat transfer in air. Simulations were performed for different orientations or tilt angles of the square cylinder spanning from $0^0$ to $90^0$ in the laminar range of Grashof number ($10 \leq Gr \leq 10^5$). The average $Nu$ for the square cylinder having tilt angle $45^0$ was found to be maximum in their study. Natural convection from a square cylinder in extensive unconfined non-Newtonian Power-law fluid has been numerically investigated by Sasmal and Chhabra [8] in laminar regime of Grashof number spanning in the range 10 to $10^5$. Computations were done for various Prandtl numbers including air ($Pr = 0.72$) and for a wide range of power-law index ($n$) encompassing both shear-thinning and shear-thickening fluid behaviors including the Newtonian case ($n =1$). A recirculation zone has been seen at the top of the square cylinder at higher $Gr$ only. The local $Nu$ was found to increase significantly near the sharp corners of the cylinder. Sasmal and Chhabra [9] numerically studied the laminar free convection from a $45^0$ tilted square cylinder in unconfined power-law fluids. For the case of Newtonian fluid ($n=1$), the formation of vortices was completely absent even at a higher $Gr$ due to the smooth flow of the buoyant plume over the tilted square. The average $Nu$ was found to increase with increase in $Gr$, whereas the drag coefficients were decreased in their study. Experimental and numerical analysis of laminar natural convection from sharp-edged horizontal bars has been conducted by Chang et al. [10]. The completely Navier-Stokes equations along with the energy equation for a square cylinder were numerically solved using a finite-difference scheme for $Pr$ =0.7. Owing to the boundary layer separation at higher Rayleigh number ($Ra$ >5×10$^3$), twin vortices were identified at the upper edge of the square cylinder. The numerical results were compared with the experimental data with good accuracy. Temperature inversions over a square cylinder were experimentally visualized in natural convection by Cho and Chang [11] at moderate Grashof number.

The aforementioned literature on natural convection is restricted to a single horizontal cylinder. However, it has been readily acknowledged that natural convection from multiple horizontal cylinders of various shapes and orientations has significant engineering applications [12,13]. There exist a limited body of literature on unconfined free convection from multiple horizontal cylinders of various orientations. Experimental and numerical study on natural convection from two vertically arranged circular cylinders have been conducted by Chae and Chung [14]. The ratio of Nusselt numbers of upper to lower cylinder was found to increase gradually with increase in pitch to diameter ratio. Cianfrini et al. [15] investigated the steady-state natural convection numerically from a heated circular cylinder in the presence of a downstream cylinder. The interaction of their thermal plumes was found to affect the overall heat transfer of the system. Experimental and numerical investigation of free convection from two vertically staggered circular cylinders have been conducted by Heo et al. [16]. Owing to the strong interaction and preheating of the thermal plume, the heat transfer from the top cylinder was reduced significantly due to the presence of the bottom cylinder. This effect is known as temperature difference imbalance in literature which was previously reported in the numerical investigation of Park and Chang [17]. Natural convection heat transfer from two vertically attached horizontal circular cylinders have been numerically investigated by Liu et al. [18]. The heat transfer from individual cylinders was decreased due to the formation of recirculation zones near the contact point of the two cylinders.

Lie et al [19] numerically studied the laminar natural convection from two horizontally attached horizontal cylinders in air. Owing to the formation of a compact body shape, vortices were developed at the above and below the contact point of the two cylinders and the size of these vortices gradually increased with increase in the Rayleigh number. The optimal spacing between horizontal cylinders corresponding to maximum natural convection heat transfer has been demonstrated experimentally, analytically and numerically by Bejan et al. [20]. Correlations for optimal spacing and maximum heat flux were also developed in their study for future correspondence. Rath and Dash [21,22] numerically investigated the natural convection heat transfer from a stack of horizontal cylinders in 2D and 3D computational domains in the respective papers. Computations were conducted in both laminar and turbulent flow regimes for three, six and ten cylinders arranged in a triangular stack. Razzaghpanah et al. [23] numerically studied the laminar natural convection from a single row of closely spaced isothermal horizontal cylinders in molten solar salt. A statistical model was used to develop a Nusselt number correlation as a function of Rayleigh number and spacing between the cylinders.

To the best of our knowledge, substantial research on natural convection heat transfer from various shapes of bluff bodies are now available in present-day literature, however natural convection from horizontally aligned square cylinders are completely absent. Hence, the present study is motivated to investigate the effect of horizontal spacing ($S/W$) and tilt angle ($\delta$) on natural convection heat transfer from two horizontally aligned square cylinders in the laminar flow regimes. The practical engineering applications of such configuration can be found in thermal heat storage technology, waste heat recovery applications, drying of food stuff, heat exchanger designs and so on. The present work invokes the computational fluid dynamics (CFD) for effective investigation of natural convection to delineate both momentum and heat transfer characteristics along with the thermodynamic performance of the system. Bejan's approach [20] is implemented to choose an economical computational domain and a GCI method [24-26] is used to justify the grid independence for the present numerical study. The numerical results represented herein are in terms of velocity and temperature profiles, isotherm contours and stream-function lines, local and average $Nu$, mass flow rate, optimum spacing, drag coefficients, entropy generations, and finally concluded with developing a suitable empirical correlation for average Nusselt number.

## 2. Problem Statement

### 2.1. Physical description

The objective of the present study is to conduct the 2-dimensional numerical simulations for external natural convection from bluff bodies in an unconfined fluid medium. Hence, two horizontally aligned isothermal square cylinders are immersed in an extensive quiescent Newtonian fluid medium and maintained at a temperature $T_w$ higher than the ambient quiescent fluid ($T_w > T_\infty$) as shown in Figure 1. The size of the square cylinders is taken as '$W$' and the horizontal spacing between the cylinders as '$S$'. The origin of the Cartesian coordinate frame ($x, y$) in the present computational domain is located exactly at the centre of the right cylinder and gravity acts along the negative $y$-direction. Owing to the presence of temperature gradient in the domain (as a result of thermal expansion), a density gradient eventuates in the vicinity of the two cylinders which leads to the appearance of an upward buoyancy motion of the fluid over the cylinders. Figure 2 shows different orientations or tilt angle ($\delta$) of the square cylinder such as $\delta=0^0$ or $90^0$, $\delta=30^0$, $\delta=45^0$, and $\delta=60^0$.

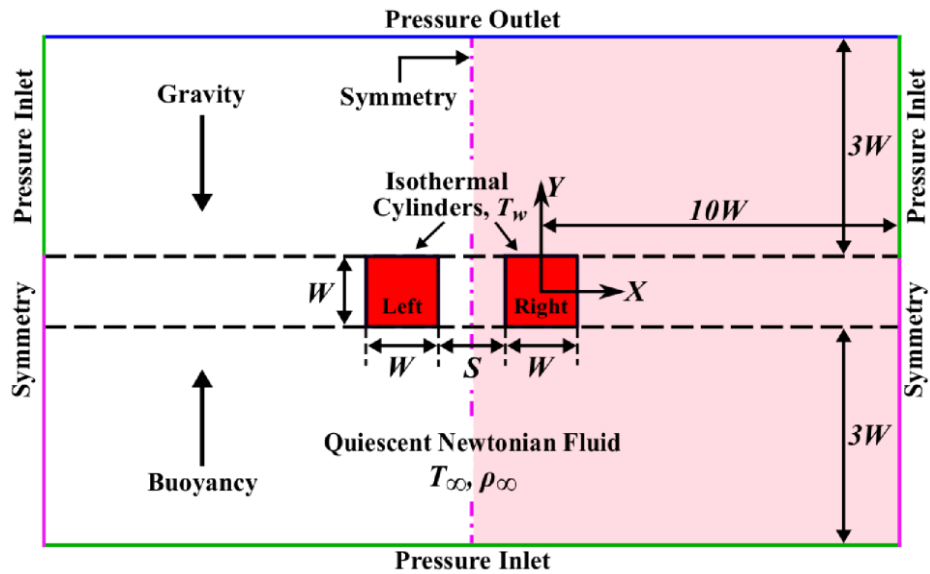

**Figure 1.** Schematic of the physical problem with boundary conditions as per Bejan's approach (dimensions are not to scale)

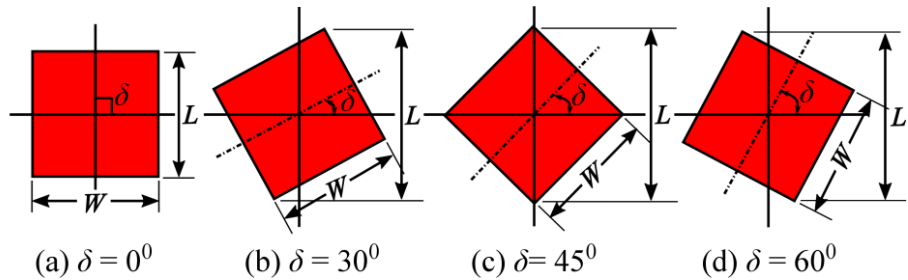

(a) $\delta = 0^0$  (b) $\delta = 30^0$  (c) $\delta = 45^0$  (d) $\delta = 60^0$

**Figure 2.** Different orientations (tilt angle) of the square cylinder

For an extensive unconfined flow problem, it is customary to immerse the bluff body in a large computational domain, like in excess of 100 times [14] or even 1200 times the size of the body [8,9] which leads to the requirement of a large number of computational cells and enormous computational resources. Hence, the purposed Bejan's approach [20] is applied in the present study to reduce the size of the domain to a great extent. This approach was also implemented subsequently by Liu et al. [18,19]. The flow is expected to be steady and symmetric about the vertical center line between the square cylinders hence, computations have been conducted only for a half (right half) of the domain and called as computational domain. According to Bejan's demonstration [20], the right face of the domain from bottom face to the top of the cylinder is also assigned as a symmetry boundary. The bottom face along with the top right face of the domain was treated as pressure inlet boundaries, whereas, the top face of the domain is designated as pressure outlet boundary as shown in Figure 1. It can be noted here that the pressure inlet boundary assigned at the top right face of the domain is just to avoid any artificial chimney effect in the domain. The details of the domain size can be seen in section 3.1 of this paper.

*2.2. Mathematical modeling*

In the present study, the thermo-buoyant flow is assumed to be two-dimensional, steady, laminar, and incompressible. The fluid is assumed to be Newtonian and all the thermo-physical properties, except the density of the fluid, are taken as constant at mean temperature ($T_m$). The temperature dependent density (the driving force for natural convection) is captured customarily by employing the well-known "Boussinesq Approximation" as Eq. (1). It assumes the density to be constant in all the governing equations except the body force term in the y-momentum equation.

$$(\rho_\infty - \rho) \approx \rho\beta(T - T_\infty) \quad (1)$$

It is worthwhile to mention here that the use of constant thermo-physical properties at mean temperature ($T_m$) is justified since the maximum temperature difference ($T_w - T_\infty$) in the present study is set to 2K. It is quite reasonable to assume the thermal expansion coefficient as a constant value as expressed in Eq. (2). Furthermore, the radiation heat transfer is ignored in this study since the temperature being very small.

$$\beta = -\frac{1}{\rho}\left(\frac{\partial \rho}{\partial T}\right)_P = \frac{1}{T_m} \quad (2)$$

*2.2.1. Governing equations*

In accordance with the aforementioned assumptions, the governing conservation equations are simplified and written in the dimensional Cartesian frame as follows:

*Continuity:*
$$\frac{\partial u}{\partial x} + \frac{\partial v}{\partial y} = 0 \quad (3)$$

*x- Momentum:*
$$u\frac{\partial u}{\partial x} + v\frac{\partial u}{\partial y} = -\frac{1}{\rho}\frac{\partial p}{\partial x} + \frac{\mu}{\rho}\left[\frac{\partial^2 u}{\partial x^2} + \frac{\partial^2 u}{\partial y^2}\right] \quad (4)$$

*y- Momentum:*
$$u\frac{\partial v}{\partial x} + v\frac{\partial v}{\partial y} = -\frac{1}{\rho}\frac{\partial p}{\partial y} + \frac{\mu}{\rho}\left[\frac{\partial^2 v}{\partial x^2} + \frac{\partial^2 v}{\partial y^2}\right] + g\beta(T - T_\infty) \quad (5)$$

*Thermal Energy:*
$$\rho C_p\left(u\frac{\partial T}{\partial x} + v\frac{\partial T}{\partial y}\right) = K\left(\frac{\partial^2 T}{\partial x^2} + \frac{\partial^2 T}{\partial y^2}\right) + \mu\Phi \quad (6)$$

All the dimensional quantities appearing in the governing Eqs. (3-6) are defined non-dimensionally as follows:

$$x^* = \frac{x}{L}, \ y^* = \frac{y}{L}, \ u^* = \frac{u}{U_c}, \ v^* = \frac{v}{U_c}, \ p^* = \frac{p}{\rho_\infty U_c^2}, \ \varphi = \frac{T - T_\infty}{T_w - T_\infty} \quad (7)$$

Where, $L = W(\cos\delta + \sin\delta)$ and $U_c = \sqrt{g\beta(T_w - T_\infty)L}$ are the reference or characteristic length scale and velocity scale, respectively.

### 2.2.2. Boundary conditions

Some suitable physically realistic boundary conditions are imposed in the computational domain as shown in Figure 1 to solve the governing conservation Eqs. (3-6), which can be written in mathematical form as follows:

*Cylinder surface:* Isothermal condition along with no slip and no penetration wall boundary conditions are imposed on the surface of the square cylinders, i.e.:

$$T = T_w, \ u = v = 0 \quad (8)$$

*Symmetry:* Planner symmetry boundary condition is assigned at the symmetry faces. i.e.:

$$u = 0, \ \frac{\partial v}{\partial x} = 0, \ \frac{\partial T}{\partial x} = 0 \quad (9)$$

*Pressure inlet:* The bottom face and the top-right face of the flow domain are designated as pressure inlet boundaries where the pressure is set to atmospheric condition (zero total gauge pressure). The temperature of the entering fluid is also set to ambient condition at this boundary.

at the bottom face: $\quad p = p_\infty = 0, \ u = \frac{\partial v}{\partial y} = 0, \ T = T_\infty \quad (10)$

at the top-right face: $\quad p = p_\infty = 0, \ v = \frac{\partial u}{\partial x} = 0, \ T = T_\infty \quad (11)$

*Pressure outlet:* A pressure outlet boundary condition is assigned at the top face of the flow domain. At this boundary, pressure is set to atmospheric pressure (zero static gauge pressure) and the temperature of any back-flow is set to be same as the surroundings quiescent fluid.

$$p = p_\infty = 0, \ u = \frac{\partial v}{\partial y} = 0, \ T = T_\infty \quad (12)$$

### 2.2.3. Calculation for entropy generation

The total rate of entropy generation is consolidated by two major thermodynamic irreversible consequences. One is the irreversible thermal diffusion due to local temperature gradient which is commonly known as entropy generation due to heat transfer, whereas the other one is the entropy generation due to the fluid friction [27-30]. Hence the rate of entropy generation per unit volume can be written as:

$$\dot{s}_{gen} = \dot{s}_{gen,HT} + \dot{s}_{gen,FF} \tag{13}$$

Where, the rate of entropy generation per unit volume due to heat transfer is defined as:

$$\dot{s}_{gen,HT} = \frac{K}{T^2}\left\{\left(\frac{\partial T}{\partial x}\right)^2 + \left(\frac{\partial T}{\partial y}\right)^2\right\} \tag{14}$$

Hence, the rate of total entropy generation due to heat transfer can be calculated as:

$$\dot{S}_{gen,HT} = \iiint_V \dot{s}_{gen,HT}\, dV \tag{15}$$

The thermodynamic irreversibility due to heat transfer ($I_{HT}$) can be defined as:

$$I_{HT} = T_\infty \dot{S}_{gen,HT} \tag{16}$$

In Eq. (13), the rate of entropy generation due to fluid friction per unit volume is defined as:

$$\dot{s}_{gen,FF} = \frac{\mu}{T}\Phi \tag{17}$$

Where, $\Phi$ is the viscous dissipation per unit volume and can be defined as:

$$\Phi = 2\left[\left(\frac{\partial v}{\partial y}\right)^2 + \left(\frac{v}{y}\right)^2 + \left(\frac{\partial u}{\partial x}\right)^2 + \frac{1}{2}\left(\frac{\partial v}{\partial x} + \frac{\partial u}{\partial y}\right)^2\right] \tag{18}$$

The rate of total entropy generation due to fluid friction can be calculated as:

$$\dot{S}_{gen,FF} = \iiint_V \dot{s}_{gen,FF}\, dV \tag{19}$$

Hence, the thermodynamic irreversibility due to fluid friction ($I_{FF}$) can be defined as:

$$I_{FF} = T_\infty \dot{S}_{gen,FF} \tag{20}$$

### 2.2.4. Dimensionless parameters

Evidently, in the present study, the numerical results are represented non-dimensionally. Hence, it is customary to explain the mathematical expression for each of those quantities.

*Grashof number* (*Gr*): It is defined as the ratio of thermo-buoyancy force to the viscous force acting on the fluid and can be written mathematically as:

$$Gr = \frac{g\beta(\Delta T)L^3}{\upsilon^2} \tag{21}$$

*Prandtl number* (*Pr*): It manifests the competing behavior between the momentum and thermal diffusions and defined as the ratio of thermal diffusivity to the momentum diffusivity.

$$Pr = \frac{\upsilon}{\alpha} = \frac{\mu C_p}{K} \tag{22}$$

*Local Nusselt number* ($Nu_l$): The convective heat flux at a location on the surface is effectively distinguished by the local Nusselt number, which can be defined mathematically as:

$$Nu_l = \frac{h_l L}{K} = -\left(\frac{\partial T}{\partial \hat{n}}\right) \tag{23}$$

*Average Nusselt number* ($Nu$): It is the area-weighted average quantity, which can be directly calculated by integrating the local values of Nusselt number over the entire surface of the cylinder.

$$Nu = \frac{hL}{K} = \frac{1}{s}\int_s Nu_l \, ds \tag{24}$$

*Drag coefficient* ($C_d$): It is the non-dimensional form of the total drag force. The resulting drag force is the sum of the normal pressure force and shearing force acting on the surface. Hence, the total drag coefficient ($C_d$) has two components such as pressure drag coefficient ($C_{pr}$) and skin friction drag coefficient ($C_{sf}$), which are defined as follows:

$$C_d = C_{pr} + C_{sf} \tag{25}$$

Where,
$$C_{pr} = \frac{F_{pr}}{1/2\rho_\infty U_c^2 L} = \int_s C_{pr,l} \, \hat{n} \, ds \tag{26}$$

The local pressure drag coefficient:
$$C_{pr,l} = \frac{P_l - P_\infty}{1/2\rho_\infty U_c^2} \tag{27}$$

and
$$C_{sf} = \frac{F_{sf}}{1/2\rho_\infty U_c^2 L} = \int_s C_{sf,l} \, \hat{n} \, ds \tag{28}$$

The local friction coefficient:
$$C_{sf,l} = \frac{\tau_l}{1/2\rho_\infty U_c^2} = \frac{2}{\sqrt{Gr}}\left(\frac{\partial U_l}{\partial \hat{n}}\right)_{wall} \tag{29}$$

*Bejan number* ($Be$): The entropy generation or thermodynamic irreversibility is defined non-dimensionally by the Bejan number [28]. It is defined as the ratio of the thermodynamic irreversibility due to heat transfer to the total irreversibility in the system.

$$Be = \frac{I_{HT}}{I_{HT} + I_{FF}} \tag{30}$$

*Mass flow rate* ($M$): The rate of mass flow in the passage between the two square cylinders are expressed non-dimensionally as follows:

$$M = \frac{\dot{m}}{\mu L} \tag{31}$$

## 3. Numerical Methodology

The present problem was solved numerically using a Finite Volume Method (FVM) approach. The computational domain was divided into a number of sub-domains commonly known as control volumes over which the governing partial differential equations (3-6) were integrated to discretize into a set of algebraic equations. The algebraic equations with the imposed boundary equations (8-12) were then solved iteratively using a pressure-based Algebraic Multi-Grid (AMG) solver available in the commercial software package ANSYS-Fluent (Academic version- 15.0.0). The SIMPLE (Semi-Implicit Method for Pressure-Linked Equations) algorithm was employed for pressure-velocity coupling, which was found to be the most stable one. Green-Gauss node based method has been used for the calculation of gradients in the discretization process. PRESTO (PREssure STaggering Option) scheme was implemented for discretization of pressure, whereas the Second Order Upwind (SOU) scheme has been applied for discretization of the convective terms. Furthermore, for a smooth convergence, the scaled residuals or the convergence criterions were set to $10^{-8}$ for energy and $10^{-6}$ for continuity and momentum equations. Convergence in the present study was checked by monitoring the average Nusselt number and drag coefficient consistent up to five significant digits. For calculation of entropy generation in the system, separate custom field functions were written in Fluent.

### *3.1. Domain independence test*

In order to make the numerical results independent of the domain size, a domain independence test has been performed over the present computational domain. In the present study, the size of the computational domain has been reduced significantly by employing the purposed Bejan's approach [20] as shown in Figure 1, where the width of the domain was taken 10 times the length scale of the problem as suggested by Bejan. Hence, Table 1 shows the independence test for the computational domain by varying the vertical height of the domain at the lowest Grashof number ($Gr$ =10) and lowest Prandtl number ($Pr$ =0.71), where the thermal boundary layer was expected to be thicker. It can be evident from Table 1 that a vertical height of 3L from bottom and top is sufficient for the present study since the change in the intended quantities ($Nu$ and $C_d$) are found to be less than 1% with further increase in the size of the domain.

**Table 1.**

Variation of average *Nu* and $C_d$ with domain height at $Gr =10$, $Pr =0.71$ for $\delta =45^0$

| Bottom height | Top height | Nu | $C_d$ |
|---|---|---|---|
| 2L | 2L | 0.8529 | 7.2303 |
| 3L | 3L | 0.8641 | 7.1201 |
| 4L | 4L | 0.8677 | 7.0682 |

*3.2. Grid independence test*

A grid independence study has been performed to make sure the numerical results remain independent of the size of the computational cells. In the present study, the boundary adaption technique was implemented to conduct the grid independence test, which avoids the unnecessary refinement of the cells far away from the wall. Ten cells from the cylinder wall have been adapted in three levels to conduct the test. Figure 3 shows the distribution of grids in the computational domain with the subsequent level of adaptions for two different tilt angle of the square cylinder. The grid independence test has been performed at the highest Grashof number ($Gr =10^5$) and highest Prandtl number ($Pr =7$) where the thermal boundary layer thickness was expected to be thinner and gradient in field properties near the wall would be higher. The variation of local *Nu* over the surface of a square cylinder having tilt angle $\delta =45^0$ with different level of boundary adaptions is shown in Figure 4. It can be seen that on the flat surfaces of the cylinder, the local *Nu* matches quite well for all the four types of cells whereas, near the corners of the cylinder, the local *Nu* remains almost same only beyond the level-1 adaption. The grid independence test has been further verified by the average quantities like average *Nu* and $C_d$ with the level of adaption as shown in Table 2. It is also evident from Table 2 that the variation of *Nu* and $C_d$ between level-2 and level-3 adaption is less than 1%. Hence, with reference to the preceding discussions, the grids having level-2 boundary adaption is considered to be grid independent for the numerical simulations.

**Table 2.**

Variation of *Nu* and $C_d$ with the level of boundary adaptions at $Gr =10^5$, $Pr =7$ for $\delta =45^0$

| Boundary Adaption | Nu | $C_d$ |
|---|---|---|
| No adaption | 16.0848 | 0.2337 |
| 1st adaption | 15.6089 | 0.2418 |
| 2nd adaption | 15.3383 | 0.2465 |
| 3rd adaption | 15.2674 | 0.2482 |

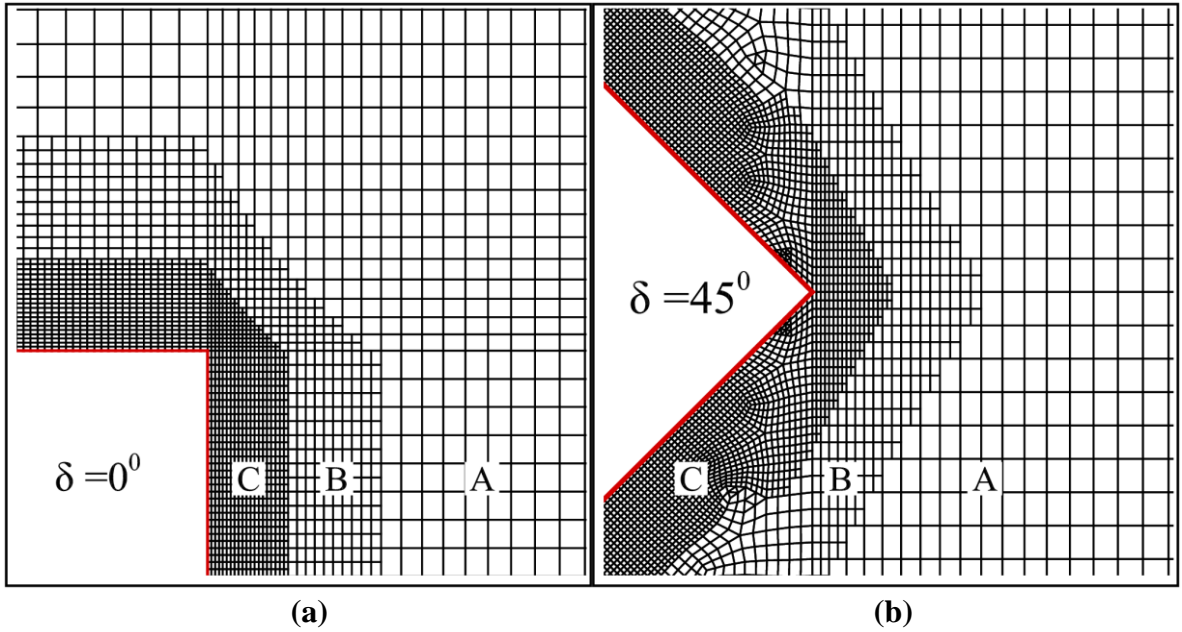

**(a)** **(b)**

**Figure 3.** Grid distribution in the computational domain with boundary adaption on the cylinder wall (A: no-adaption, B: 1$^{st}$-adaption, and C: 2$^{nd}$-adaption); for (a) square cylinder ($\delta = 0^0$) and (b) tilted square cylinder ($\delta = 45^0$)

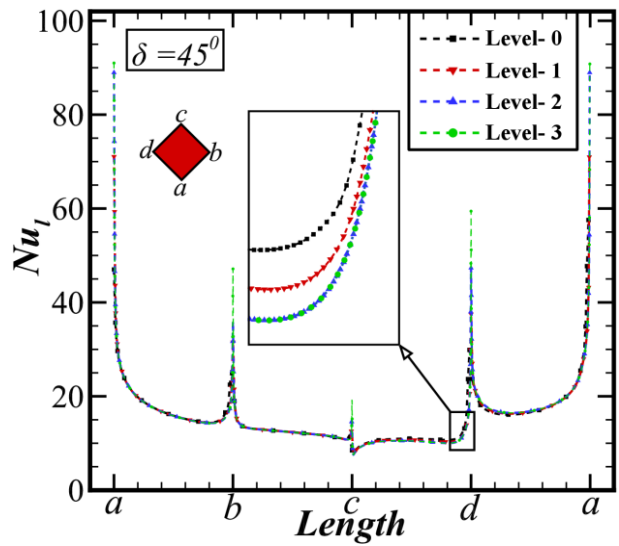

**Figure 4.** Variation of local Nusselt number ($Nu_l$) with the level of boundary adaption for a tilted square cylinder ($\delta = 45^0$) at $S/W = 1$, $Gr = 10^5$, and $Pr = 7$

### 3.3. Grid convergence index (GCI)

In the present study, a grid convergence index (GCI) has also been performed to support the grid independence test with more accuracy. The GCI algorithm is based on the Richardson extrapolation (RE) method [24,25]. Computations having lower GCI value indicates lower discretization error and better accuracy in the result. A CFD simulation having less than 1% GCI is recommended as grid independent. Table 3 shows the GCI results for a particular case having $\delta = 45^0$, $S/W = 1$, $Gr = 10^5$, and $Pr = 7$. In the present computations, the grid convergence index between level-1 and level-2 (GCI$^{1-2}$) for $Nu$ shows <1%, but it is higher for $C_d$, whereas GCI$^{2-3}$ shows less than 1% for both $Nu$ and $C_d$. Hence, the level-2 grid is suggested to take for the numerical simulations which support our customary method of grid independence study.

**Table 3.**

Grid Convergence Index (GCI) parameters

| Parameters | GCI for $Nu$ | GCI for $C_d$ |
|---|---|---|
| $N_0$ | 62813 | 62813 |
| $N_1$ | 77117 | 77117 |
| $N_2$ | 106052 | 106052 |
| $N_3$ | 164426 | 164426 |
| $r_{01}$ | 2.0459 | 2.0459 |
| $r_{12}$ | 2.0383 | 2.0383 |
| $r_{23}$ | 2.0148 | 2.0148 |
| $\phi_0$ | 15.3975 | 0.2401 |
| $\phi_1$ | 15.3624 | 0.2444 |
| $\phi_2$ | 15.3383 | 0.2465 |
| $\phi_3$ | 15.3434 | 0.2467 |
| $\varepsilon^{0-2}$ | 0.5192 | 1.0259 |
| $\varepsilon^{1-3}$ | 2.1870 | 3.0190 |
| $\phi_{ext}^{1-2}$ | 15.2844 | 0.2484 |
| $\phi_{ext}^{2-3}$ | 15.3448 | 0.2467 |
| $e_a^{1-2}$ | 0.0016 | 0.0084 |
| $e_a^{2-3}$ | 3.3239e-04 | 9.7276e-04 |
| $e_{ext}^{1-2}$ | 0.0035 | 0.0077 |
| $e_{ext}^{2-3}$ | 9.1619e-05 | 1.3344e-04 |
| $GCI^{1-2}$ | 0.4390% | 1.0975% |
| $GCI^{2-3}$ | 0.01145% | 0.0167% |

## 4. Result and Discussion

In the present study, the natural convection heat transfer from two horizontally aligned square cylinders in different Newtonian fluids are numerically investigated to elucidate both momentum and heat transfer characteristics under the following pertinent parameters: Grashof number ($10 \leq Gr \leq 10^5$), Prandtl number ($0.71 \leq Pr \leq 7$), horizontal spacing ($0 \leq S/W \leq 10$), and tilt angle ($0 \leq \delta \leq 60^0$). Hence, total of 360 numerical simulations were performed using 4-core parallel processing in Intel- core i7 processor having 3.4 GHz clock-speed. For the sake of brevity, the numerical results are represented only for some selected parameters in terms of velocity and temperature profiles, temperature contours and stream function lines, local and average Nusselt number, the mass-flow rate in the passage between the two cylinders, and the drag coefficient values. This paper is finally concluded with the development of an empirical correlation for the average Nusselt number. However, before reproducing the numerical results, it is customary to validate the accuracy and reliability of the implemented numerical methodology with prior literature.

*4.1. Model validation*

The present numerical schemes are validated with the existing experimental and numerical results for a single square cylinder. Some special simulations are performed for this purpose to compare with the existing results. Our numerical results for an unconfined single square cylinder ($\delta =0^0$) in air ($Pr =0.71$) are compared with the numerical results of Sasmal and Chhabra [10], Mahmud et al. [7], and experimental results of Eckert and Soehngen [5]. Figure 5 shows the distribution of local Nusselt number ($Nu_l$) over the surface of the cylinder for various flow conditions. The present numerical results in Figure 5(a) show an excellent agreement with the numerical results of Sasmal and Chhabra [8] at three different *Gr*. It also shows a close resemblance with Eckert and Soehngen [5] and Sasmal and Chhabra [8] at Ra $=4\times10^4$ as shown in Figure 5(b). The average *Nu* and $C_d$ for a single square cylinder are also compared in Table 4 and it shows a quite good agreement with the Sasmal & Chhabra [8] with less than 2% deviation, Mahmud et al. [7] with less than 6% deviation, and Chang et al. [10] with less than 3% deviation.

The numerical schemes are further validated with Sasmal and Chhabra [9] and Mahmud et al. [7] for a single tilted square cylinder ($\delta =45^0$) to gain more confidence. Figure 6 shows the comparison of local Nusselt number ($Nu_l$) with the existing results at two different *Gr*. The present results show an excellent agreement with Sasmal and Chhabra [9], whereas it is almost satisfactory with Mahmud et al. [7]. Table 5 shows a good agreement of average *Nu* and $C_d$ with the Sasmal and Chhabra [9] with less than 3% deviation and Mahmud et al. [7] with less than 7% deviation. Hence, with the aforementioned comparisons, we got enough confidence in our numerical schemes which would provide much accurate and reliable results for the present study.

**Table 4.**
Comparison of present results for a square cylinder ($\delta =0^0$) in air ($Pr =0.71$) with Sasmal & Chhabra [8], Mahmud et al. [7], and Chang et al. [10].

| Ra | Nu | | Gr | Nu | | | $C_d$ | |
|---|---|---|---|---|---|---|---|---|
| | Present | Ref. [10] | | Present | Ref. [8] | Ref. [7] | Present | Ref. [8] |
| $4\times10^4$ | 5.1728 | 5.3635 | 10 | 1.0589 | 1.0803 | 1.1302 | 11.1278 | 10.8048 |
| $5\times10^4$ | 5.4851 | 5.6225 | $10^3$ | 2.4431 | 2.4867 | 2.5817 | 2.6327 | 2.5618 |
| $1\times10^5$ | 6.5386 | 6.6863 | $10^5$ | 6.2557 | 6.3492 | 6.6037 | 0.9762 | 0.9509 |

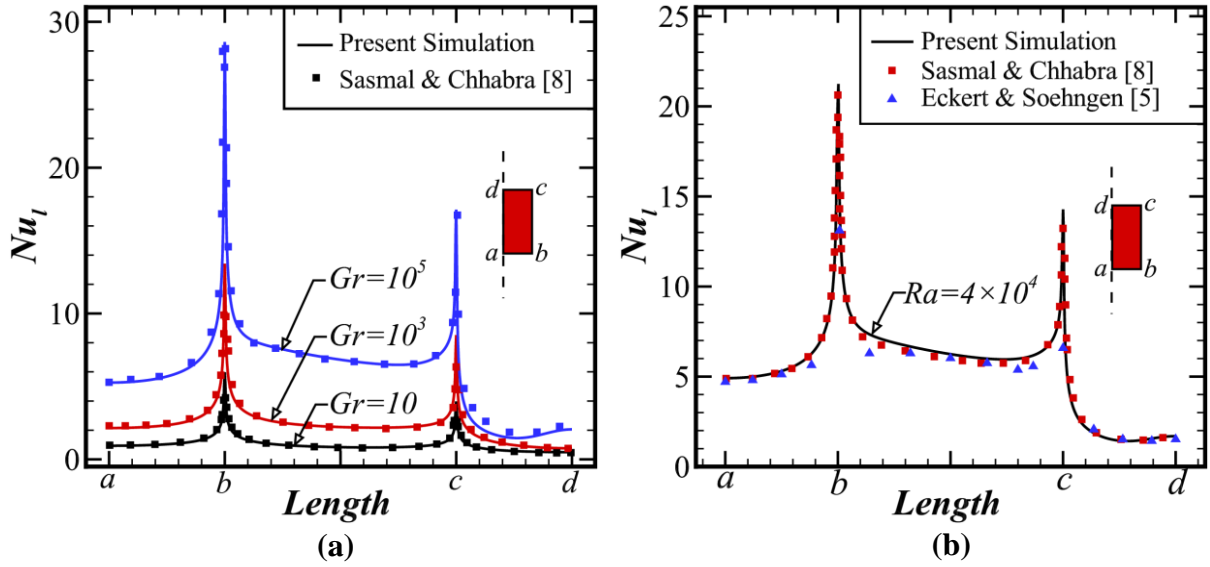

**Figure 5.** Comparison of present local $Nu$ for a square cylinder ($\delta =0^0$) in air ($Pr =0.71$) with (a) Sasmal & Chhabra [8] as a function of $Gr$ and (b) Sasmal & Chhabra [8] and Eckert & Soehngen [5] at $Ra =4\times10^4$.

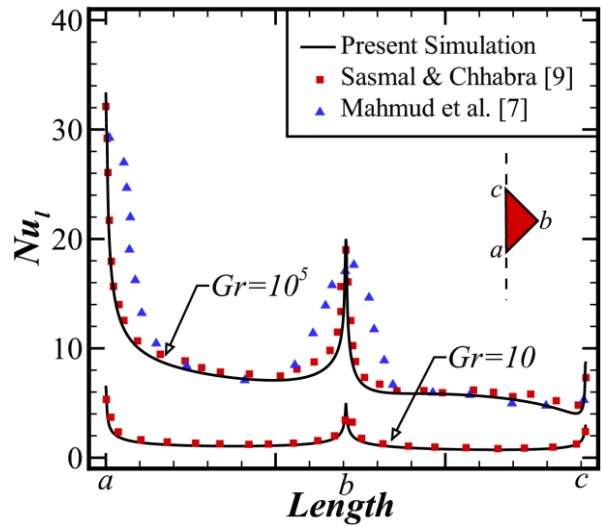

**Figure 6.** Comparison of present local $Nu$ for a tilted square cylinder ($\delta =45^0$) in air ($Pr =0.71$) with Sasmal & Chhabra [9] and Mahmud et al. [7] as a function of $Gr$

**Table 5.**
Comparison of present results for a square cylinder ($\delta =45^0$) in air ($Pr =0.71$) with Sasmal & Chhabra [9], Mahmud et al. [7].

| $Gr$ | $Nu$ | | | $C_d$ | |
|---|---|---|---|---|---|
| | Present | Ref. [9] | Ref. [7] | Present | Ref. [9] |
| 10 | 1.1081 | 1.1384 | 1.1808 | 8.3873 | 8.1023 |
| $10^3$ | 2.8159 | 2.8889 | 2.9735 | 1.6737 | 1.6145 |
| $10^5$ | 7.7308 | 7.9389 | 8.1246 | 0.4328 | 0.4211 |

## 4.2. Temperature contours and stream function lines

In the present study, both thermal and flow fields are qualitatively visualized in terms of temperature contours and streamlines, respectively. Owing to steady and symmetric buoyant flow about the vertical center line between the cylinders, Figure 7-10 show the normalized temperature contours in the left half and non-dimensional stream function lines in the right half of the computational domain for some pertinent conditions. The maximum stream function value for each case is also shown for a better understanding of the flow field quantitatively. It can be seen that with increase in Grashof number ($Gr$), the strength of the thermo-buoyant flow increases and as a result the thermal boundary layer thickness gradually becoming thinner. Furthermore, with increase in Prandtl number ($Pr$), the diffusion of momentum increase and a thinner thermal boundary layer develops over the heated cylinders. Hence, it can be anticipated from the preceding discussion that the heat transfer would show a positive dependence on both $Gr$ and $Pr$.

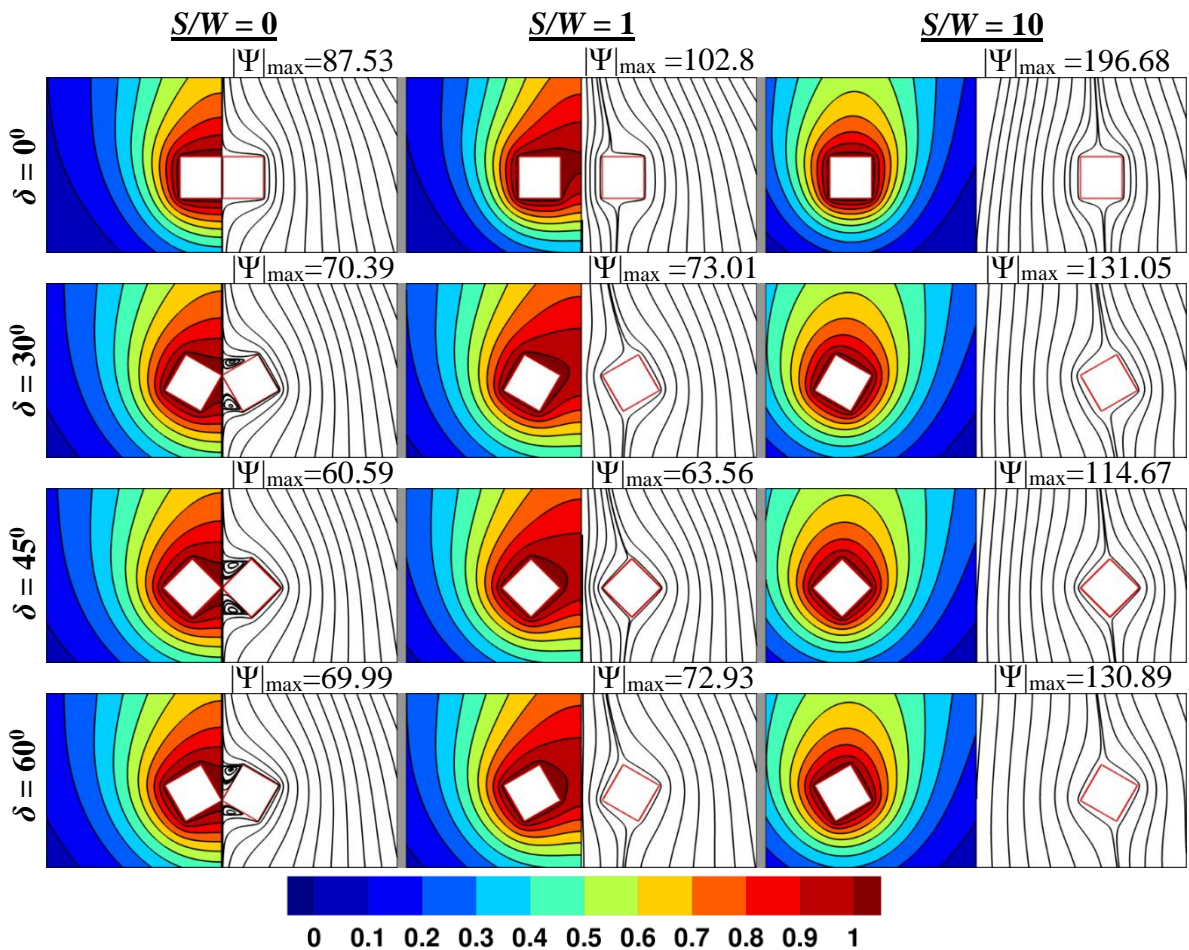

**Figure 7.** Normalized temperature contours with isotherm lines (left half) and non-dimensional stream function lines (right half) at $Gr = 10$ and $Pr = 0.71$

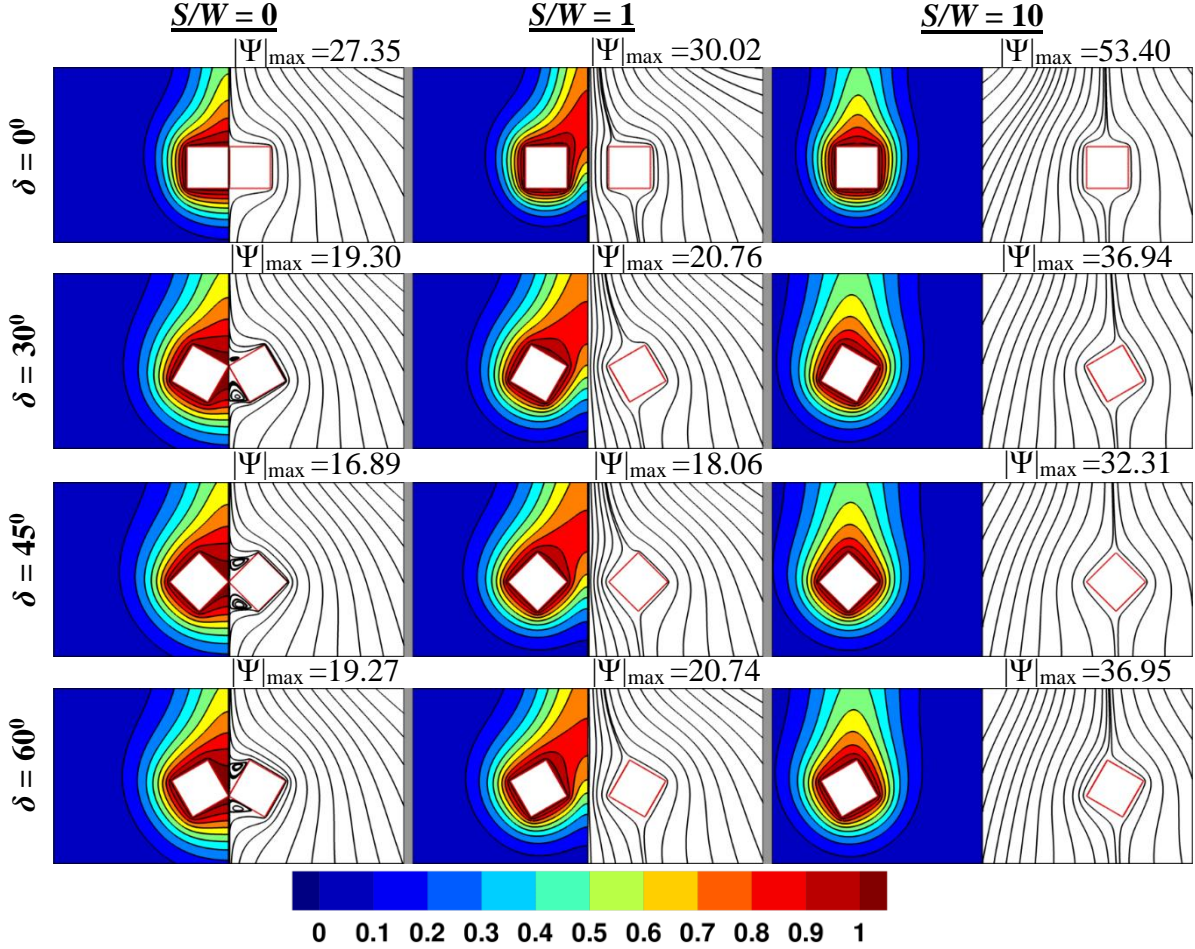

**Figure 8.** Normalized temperature contours with isotherm lines (left half) and non-dimensional stream function lines (right half) at $Gr = 10$ and $Pr = 7$

It is evident from figures that the streamline patterns show differently albeit the temperature contours are indistinguishable for different tilt angles ($\delta$) of the square cylinders. The buoyant flow tries to negotiate the shape of the bluff body and hence, bending of iso-lines are seen near the sharp corners of the square cylinders. For the case of attached cylinders ($S/W$=0), twin-vortices are formed just above and below the contact point of the two cylinders except at $\delta=0^0$. At higher $Gr$, the strength of buoyancy increase and more quiescent fluid enters from the sides of the domain hence, the size of these vortices grows with increase in $Gr$. At very high Grashof number ($Gr = 10^5$) and low Prandtl number ($Pr = 0.71$), owing to the stronger momentum and slower diffusion of flow, some small vortices are also seen above the cylinders for $\delta=30^0$ and $45^0$ at $S/W$=1 and for $\delta=0^0$ at $S/W$=10. However, due to higher momentum of the thermo-buoyant flow at higher $Gr$, these small vortices can affect the thermal field up to a certain extent. Hence, temperature inversion in the thermal fields is seen at these vortex locations as shown in Figure 9 and 10. At low horizontal spacing ($S/W\approx1$), owing to the formation of a narrow passage or chimney effect, the momentum of the flow between the cylinder passage increases and as a result the overall heat transfer increases. Somewhat straight streamlines are seen in these passage instead of bend lines and this is more prominent at high *Gr*.

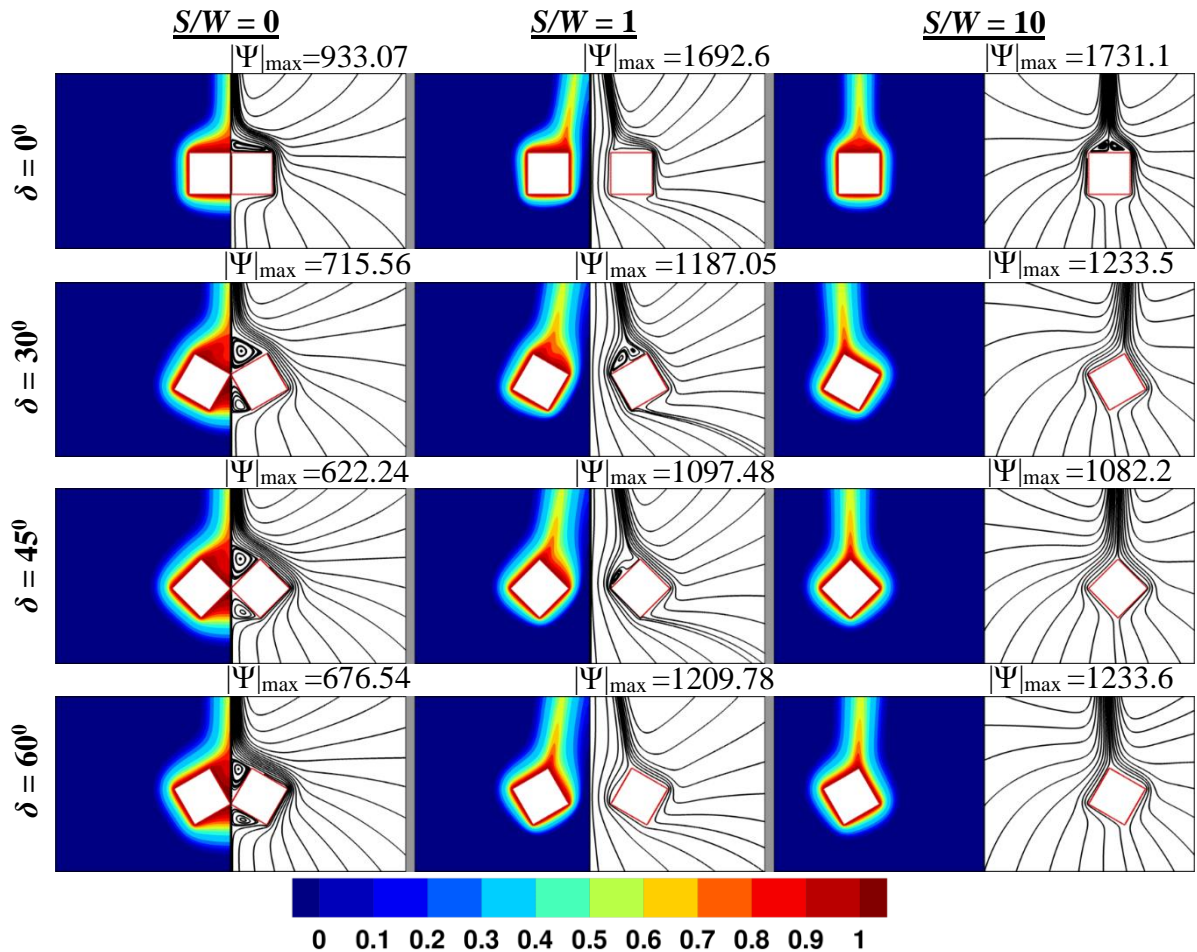

**Figure 9.** Normalized temperature contours (left half) and non-dimensional stream function lines (right half) at $Gr = 10^5$ and $Pr = 0.71$

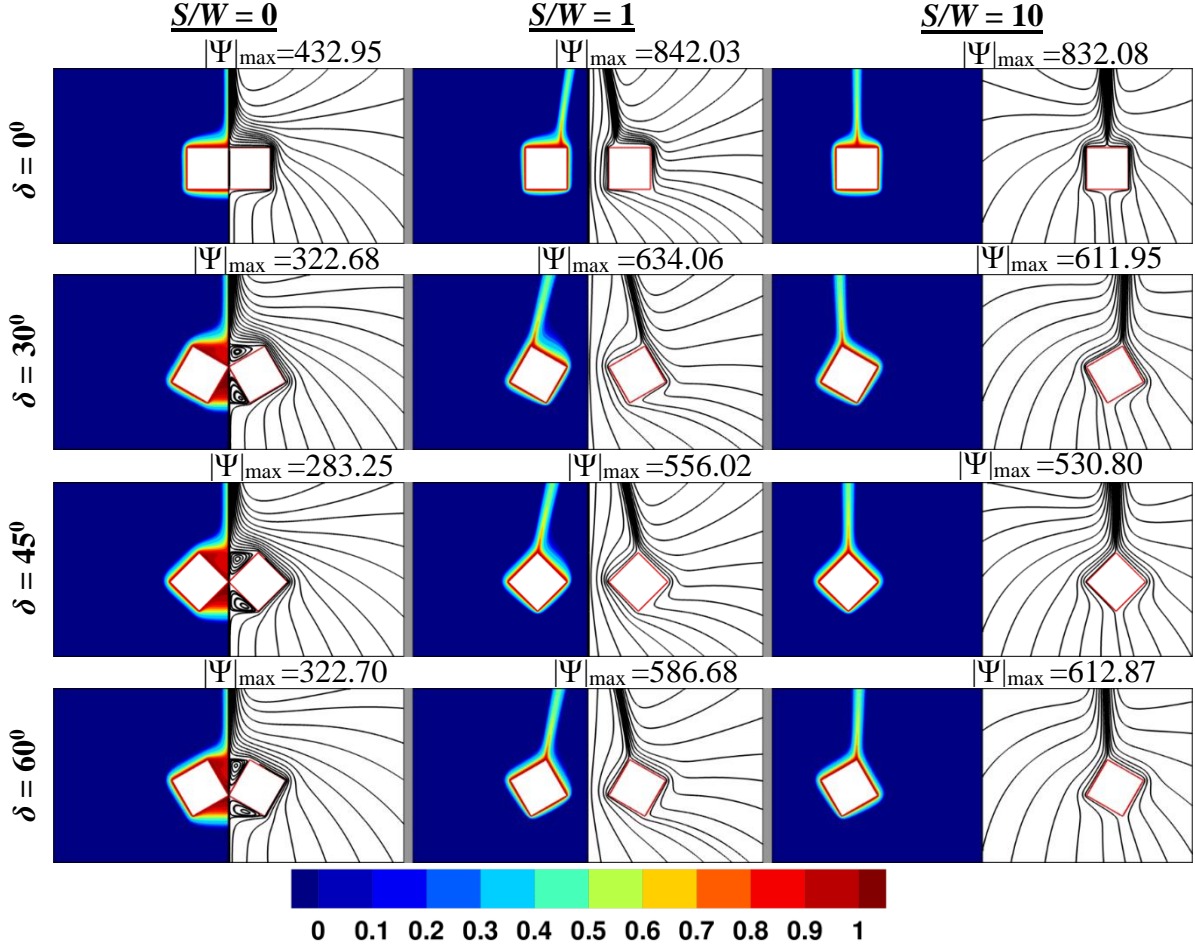

**Figure 10.** Normalized temperature contours (left half) and non-dimensional stream function lines (right half) at $Gr = 10^5$ and $Pr = 7$

### 4.3. Velocity and Temperature profiles

The momentum and thermal fields are further elucidated quantitatively by the velocity and temperature distributions along different radial directions from the cylinder surface. Figure 11 and 12 show the dimensionless velocity magnitude and normalized temperature profiles, respectively along four radial directions from the surface of a cylinder (right cylinder) for some pertinent conditions used in the present study. These profiles can reveal the boundary layer phenomenon up to a certain extent in the present study. At $\theta=0^0$, both velocity and temperature profiles are qualitatively similar to the classical natural convection profiles however, due to the restricted computational domain at other radial directions, the profiles are incomplete. It can be noticed that owing to stronger momentum of the flow, both velocity and temperature profiles are quite steeper at higher Grashof number ($Gr =10^5$) which indicates the formation of thinner boundary layer thickness and enhancement of heat transfer. At higher Prandtl number (Pr =7), both the velocity and temperature profiles fall below the profiles at low *Pr*. Hence, a thinner boundary layer thickness can be anticipated at higher *Pr*. Overall, a negative dependence on both *Gr* and *Pr* for the boundary layer thickness are observed which is indistinguishable to a single cylinder. It can also be seen from the temperature profiles in Figure 12 that the temperature of the fluid reaches the free stream temperature ($T_\infty$) much faster at $\theta=0^0$ for *S/W*=0 and at $\theta=270^0$ for *S/W* >0. Hence, heat transfer is expected to be higher at these locations.

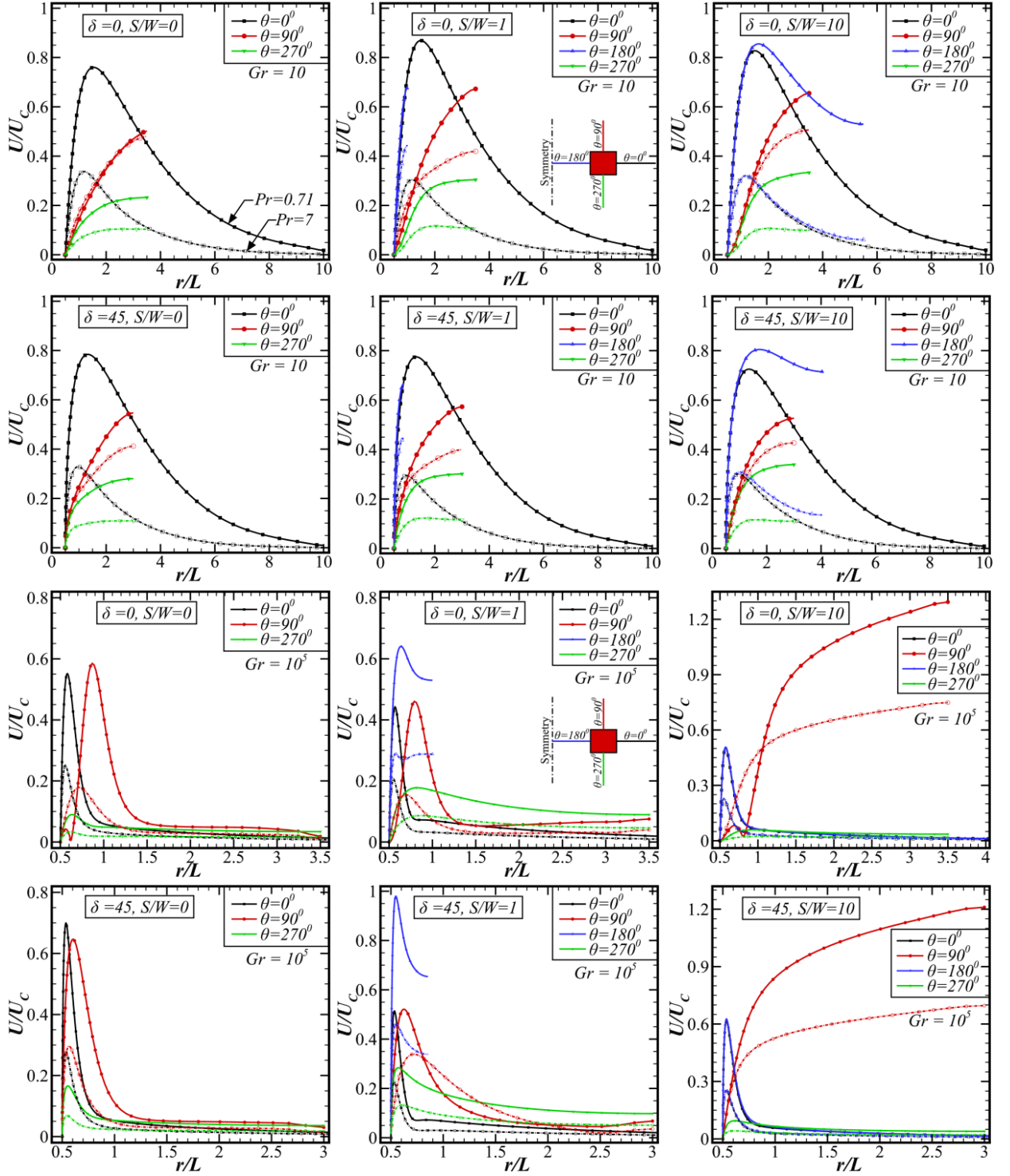

**Figure 11.** Dimensionless velocity magnitude profiles along various radial directions from the cylinder surface as a function of tilt angle ($\delta$), *S/W*, *Gr*, and *Pr* (solid line: *Pr* =0.71 and dashed line: *Pr* =7)

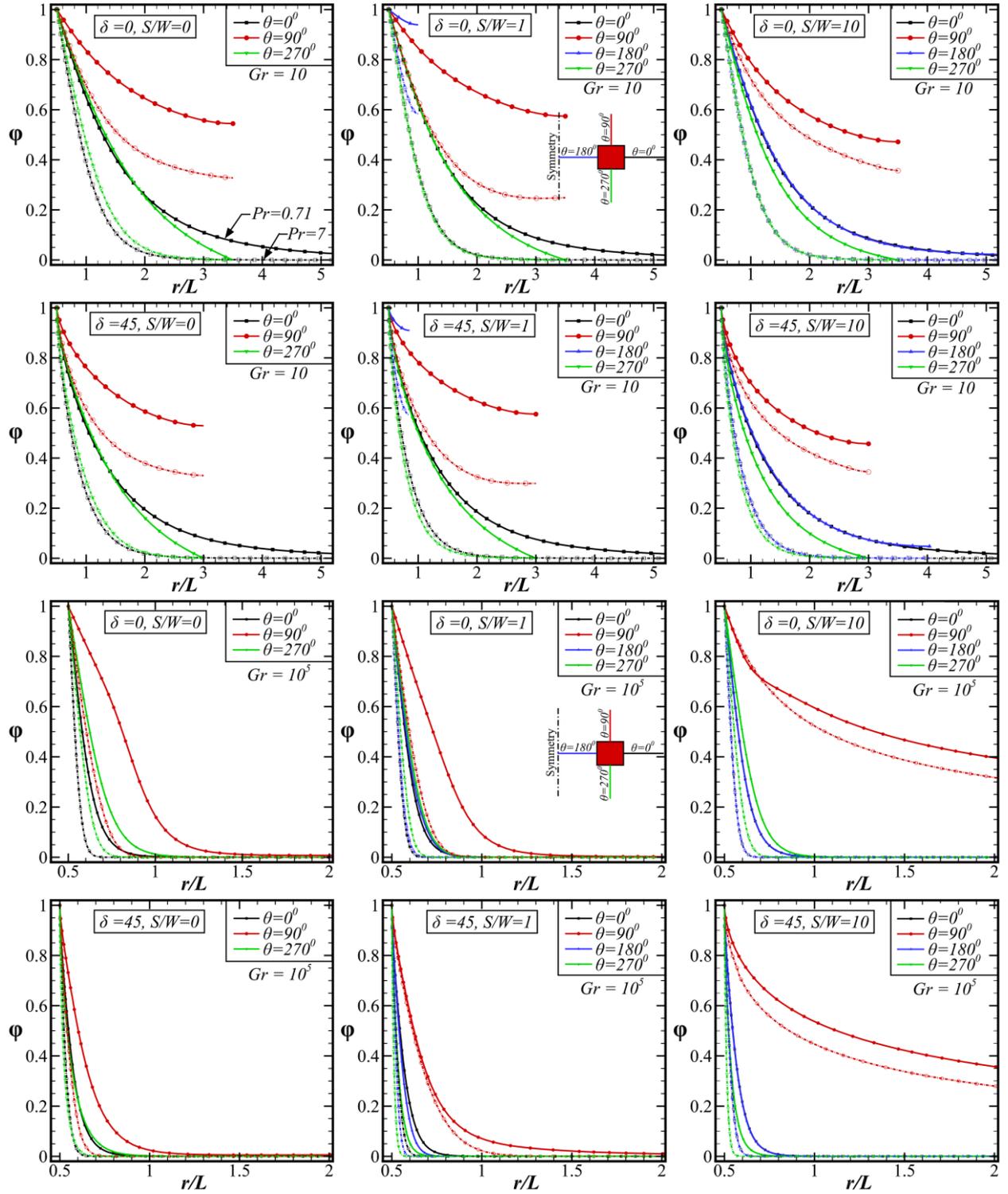

**Figure 12.** Temperature profiles along various radial directions from the cylinder surface as a function of tilt angle ($\delta$), *S/W*, *Gr*, and *Pr* (solid line: *Pr* =0.71 and dashed line: *Pr* =7)

*4.4. Mass flow rate*

In the present study, an attempt has been made to explicate the development of a chimney effect by quantifying the rate of mass flow through the passage between the two square cylinders. Figure 13 shows the non-dimensional mass flow rate with respect to the horizontal spacing (S/W) as a function of *Gr*, *Pr*, and *δ*. It is evident from Figure 13 that with an increase in Gr, the strength of the thermo-buoyant flow enhances and as a result the mass flow rate significantly increases irrespective of any other parameters. Owing to the development of a narrow passage or chimney effect between the two cylinders, the mass flow rate increases significantly and this effect is seen to be more prominent at higher *Gr*. With increase in *Pr*, the fluid behaves like more viscous which decelerate the momentum of flow and as a result the mass flow rate decreases. In the present study, the maximum non-dimensional mass flow rate is seen at low spacing (S/D≈2) for high Grashof number ($Gr=10^5$) which gradually increases with decrease in *Gr*. Owing to the non-dimensional scheme of mass flow rate, the mass flow rate is higher for the square cylinders having tilt angle $δ=0^0$ and it is lower for $δ=45^0$.

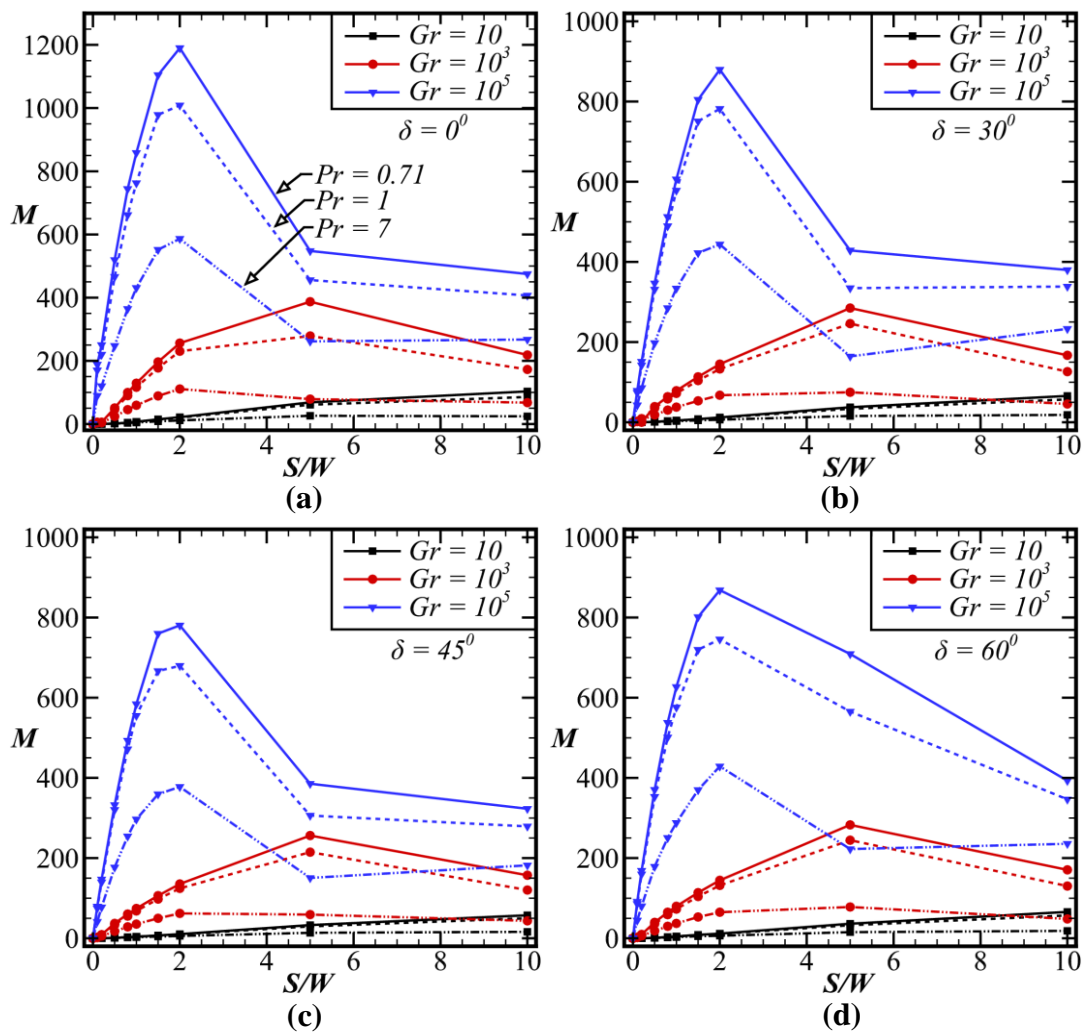

**Figure 13.** Mass flow rate through the passage between the cylinders as a function of *S/W*, *Gr*, and *Pr* (solid line: *Pr* =0.71, dashed line: *Pr* =1, dash-dot line: *Pr* =7) for (a) $δ=0^0$, (b) $δ=30^0$, (c) $δ=45^0$, and (d) $δ=60^0$

*4.5. Local Nusselt number distribution*

The local Nusselt number ($Nu_l$) can provide further physical insights into the local heat transfer characteristics over the heated surface. The local value of *Nu* is directly associated with the local temperature gradient normal to the cylinder surface which in turn associated with the velocity gradient at the surface as well as the temperature difference between the heated solid surface and the adjoining fluid. Owing to steady and symmetric flow about the vertical centreline of the domain, the representative results of the local *Nu* are explained only for the right cylinder as shown in Figure 14. For the sake of brevity, only a few extreme pertinent conditions are represented here in this article. As anticipated from the previous discussions, the local *Nu* increases with increase in *Gr* and *Pr* irrespective of horizontal spacing (*S/D*) and tilt angles (*δ*). For *S/D*=0 (attached cylinders), the thermo-buoyant flow is restricted to pass between the cylinders and hence, the heat transfer is found to be zero on the section '*d-a*' for $\delta=0^0$ and at the location' for $\delta=30^0, 45^0$, and $60^0$.

Furthermore, the buoyant plume tries to negotiate the shape of the bluff body and hence, the velocity gradients are expected to be maximum at the sharp corners (*a*, *b*, *c*, and *d*) of the square cylinder. This leads to increase in the local temperature gradient at these locations which result in increase in the local *Nu* as shown in Figure 14. Owing to the maximum local temperature gradient, the local *Nu* shows a maximum value at the location '*b*' for $\delta=0^0$, whereas it is maximum at the location '*a*' for other tilt angles. The local *Nu* gradually decreases from one sharp corner location and increases to the other corner location over the cylinder surfaces due to the gradual decrease in the velocity gradient over the surface of the cylinder. A sudden abrupt change in the local *Nu* is observed on the top surface (*c-d*) of the cylinder at high *Gr* due to the temperature inversion in the thermal field. This effect is due to the formation of vortices at higher buoyant momentum of the flow which affects the temperature contours as already discussed in the preceding section 4.2. Overall, the maximum value of local *Nu* is higher at $\delta=45^0$ compared to $\delta=60^0$ followed by $\delta=30^0$ and $0^0$, respectively.

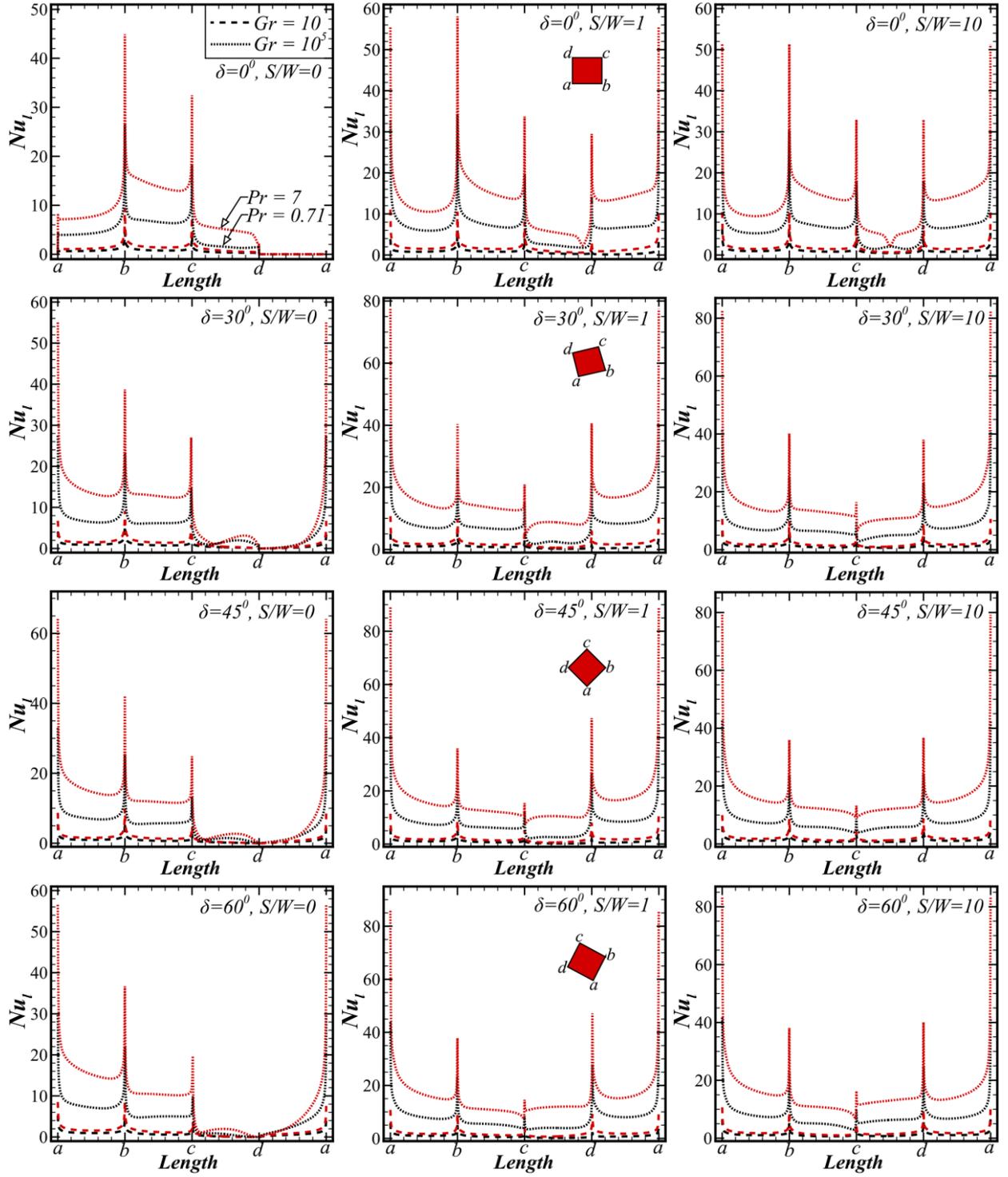

**Figure 14.** Variation of local *Nu* over the square cylinder as a function of tilt angle (*δ*), *S/W*, *Gr* (dashed line: *Gr* =10, dot line: *Gr* =$10^5$), and *Pr* (black line: *Pr* =0.71 and red line: *Pr* =7)

*4.6. Average Nusselt number*

It is customary to elucidate the average Nusselt number (*Nu*) to figure out the average heat transfer characteristics non-dimensionally which has prodigious significance in engineering calculations. Figure 15 shows the variation of average *Nu* as a function of horizontal spacing (*S/D*) for different pertinent conditions. As predicted from the previous discussions, the average surface *Nu* shows a positive dependence on both *Gr* and *Pr*. However, this dependence on *Gr* is more prominent compared to that of *Pr*. The effect of horizontal spacing (*S/D*) plays a significant role on the average *Nu*. Owing to the development of a narrow passage or chimney effect between the two horizontal square cylinders, the *Nu* increases with decrease in *S/D* up to a certain limit and the optimum horizontal spacing (*S/D$_{opt}$*) reaches corresponding to the maximum heat transfer. The optimum spacing (*S/D$_{opt}$*) corresponding to the maximum *Nu* is presented as a function of *Gr*, *Pr*, and *δ* as shown in Figure 17. It is clearly evident from Figure 17 that the *S/D$_{opt}$* shows a negative dependence on both *Gr* and *Pr*. With increase in *Gr* and *Pr*, the momentum of the thermo-buoyancy becomes stronger which leads to attain the *S/D$_{opt}$* at a smaller spacing. The tilt angle of the square cylinders shows a significant quantitative effect on the average *Nu*, albeit it is qualitatively indistinguishable. Figure 16 shows the dependence of average *Nu* with *Gr* as a function of tilt angles (*δ*). It can be seen that the average *Nu* is higher for the square cylinders having tilt angle *δ*=$45^0$ compared to *δ*=$60^0$ followed by *δ*=$30^0$ and it is minimum for *δ*=$0^0$ irrespective of *Gr*, *Pr*, and *S/W*. Furthermore, this difference in *Nu* for different tilt angles is seen to be more prominent at higher *Gr* and *Pr*. At *Gr* =$10^5$ and *Pr* =7, the average *Nu* for *δ*=$45^0$ is in excess of 22%, 3%, and 1% compared to the value at *δ*=$0^0$, *δ*=$30^0$, and *δ*=$60^0$, respectively as shown in Figure 16.

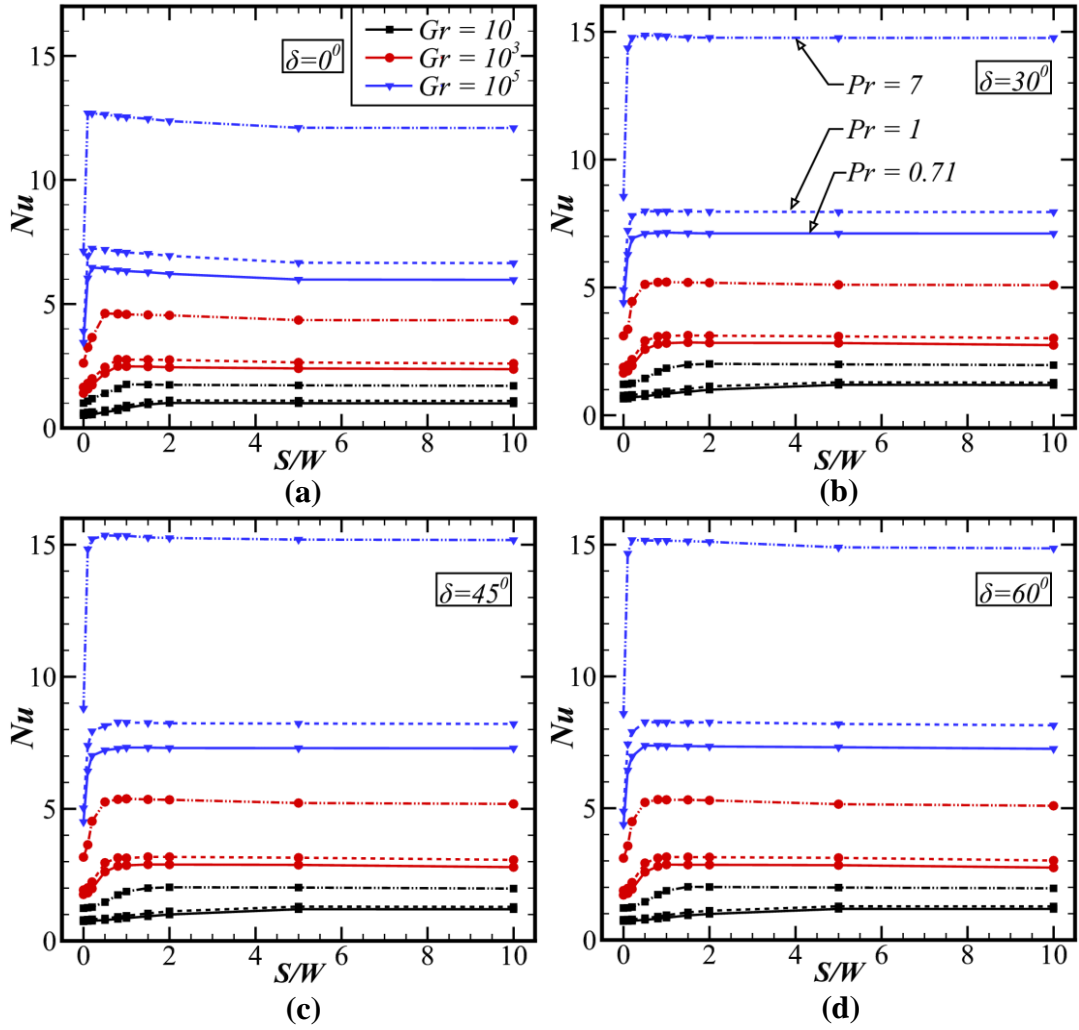

**Figure 15.** Average *Nu* as a function of *S/W*, *Gr*, and *Pr* (solid line: *Pr* =0.71, dashed line: *Pr* =1, dash-dot line: *Pr* =7) for (a) $\delta=0^0$, (b) $\delta=30^0$, (c) $\delta=45^0$, and (d) $\delta=60^0$

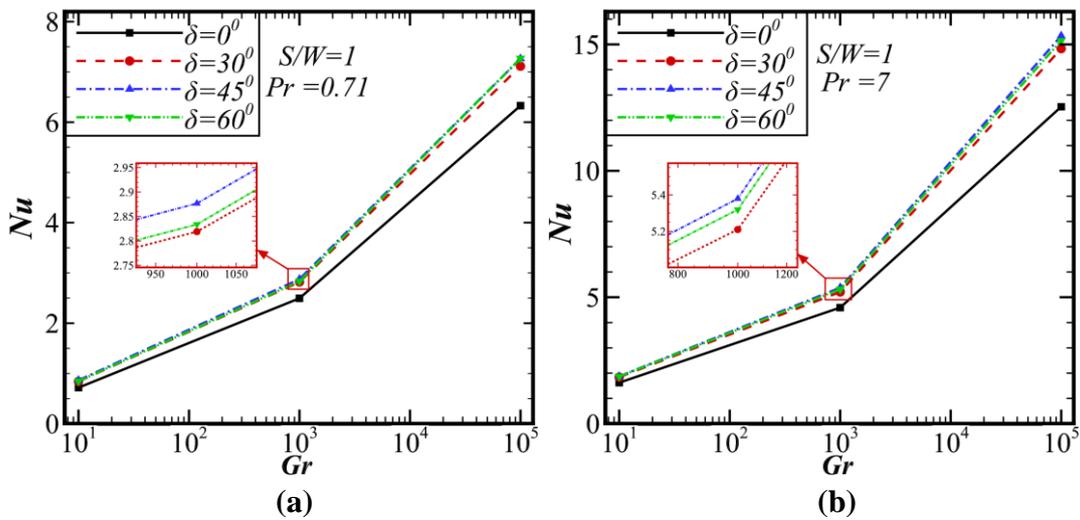

**Figure 16.** Variation of average *Nu* with *Gr* for different tilt angles (*δ*) of the square cylinder at *S/W*=1 for (a) *Pr* =0.71 and (b) *Pr* =7

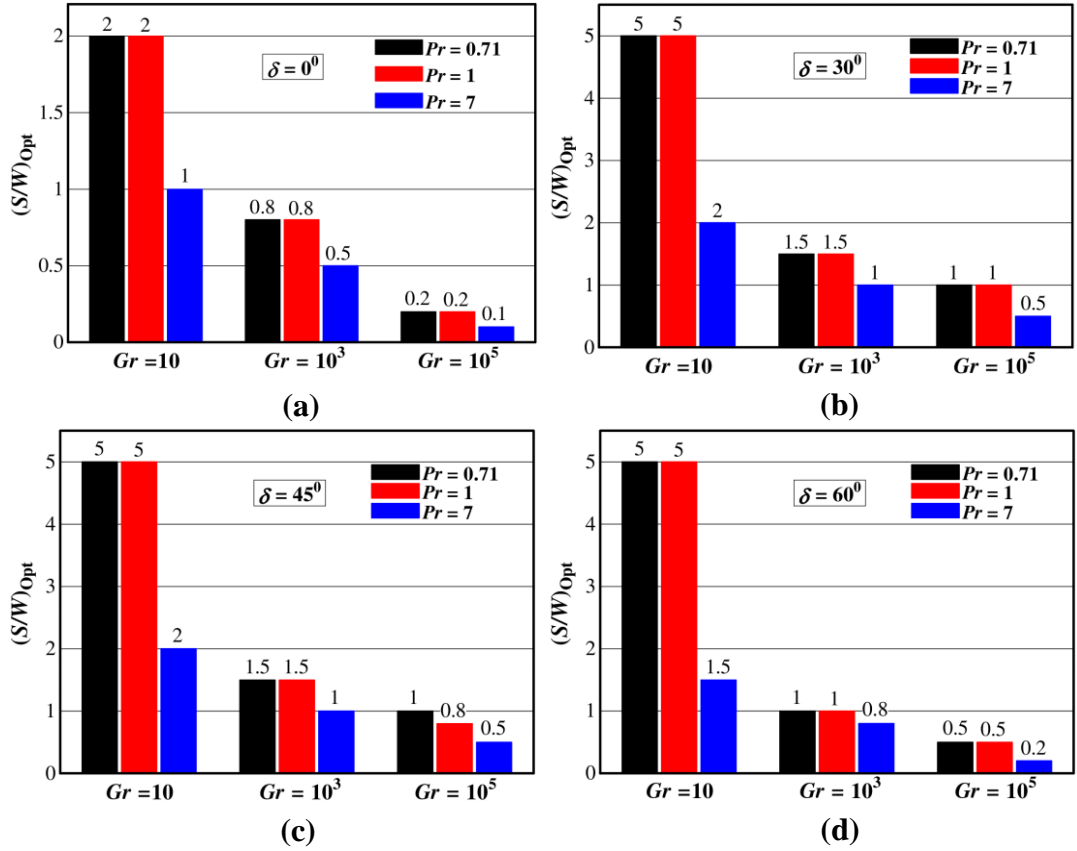

**Figure 17.** Optimum *S/W* corresponding to maximum heat transfer as a function of *Gr*, and *Pr* for (a) $\delta=0^0$, (b) $\delta=30^0$, (c) $\delta=45^0$, and (d) $\delta=60^0$

## 4.7. Entropy Generation

In the present study, though the entropy generation due to heat transfer and fluid flow in laminar natural convection are insignificant, the effect of various pertinent parameters on the entropy generation can reveal some thermodynamic aspects. Hence, an attempt is made to expound the generation of entropy in the system non-dimensionally by Bejan number (*Be*) as shown in Figure 18. The Bejan number, which is the ratio of entropy generation due to heat transfer to the total entropy generation as expressed in eqn. 30 is represented as a function of horizontal spacing (*S/W*) for various parameters like *Gr*, *Pr*, and *δ*. In the underlying physics of laminar natural convection, the momentum of the thermo-buoyant flow is quite low and as a result, the entropy generation due to fluid friction is negligible. Hence, *Be* is almost unity (100% due to heat transfer) at very low Grashof number (*Gr* =10). However, with increase in *Gr*, the fluid friction increases and ultimately the value of *Be* decreases. At low *Pr*, the fluid is slightly less viscous, which leads to increases the fluid friction. Hence, *Be* is higher at high *Pr* and decreases with decrease in *Pr*. Owing to the formation of a narrow passage or chimney effect, the momentum of the flow increase and as a result, the entropy generation due to fluid friction also increases. Therefore, *Be* is marginally lower at the optimum horizontal spacing ($S/W_{opt}$) as shown in Figure 18. The generation of entropy qualitatively shows a similar pattern for different tilt angles ($\delta$) of the square cylinder. However, due to higher heat transfer at $\delta=45^0$ (described in section 4.6), the entropy generation due to heat transfer increases which leads to a higher *Be*.

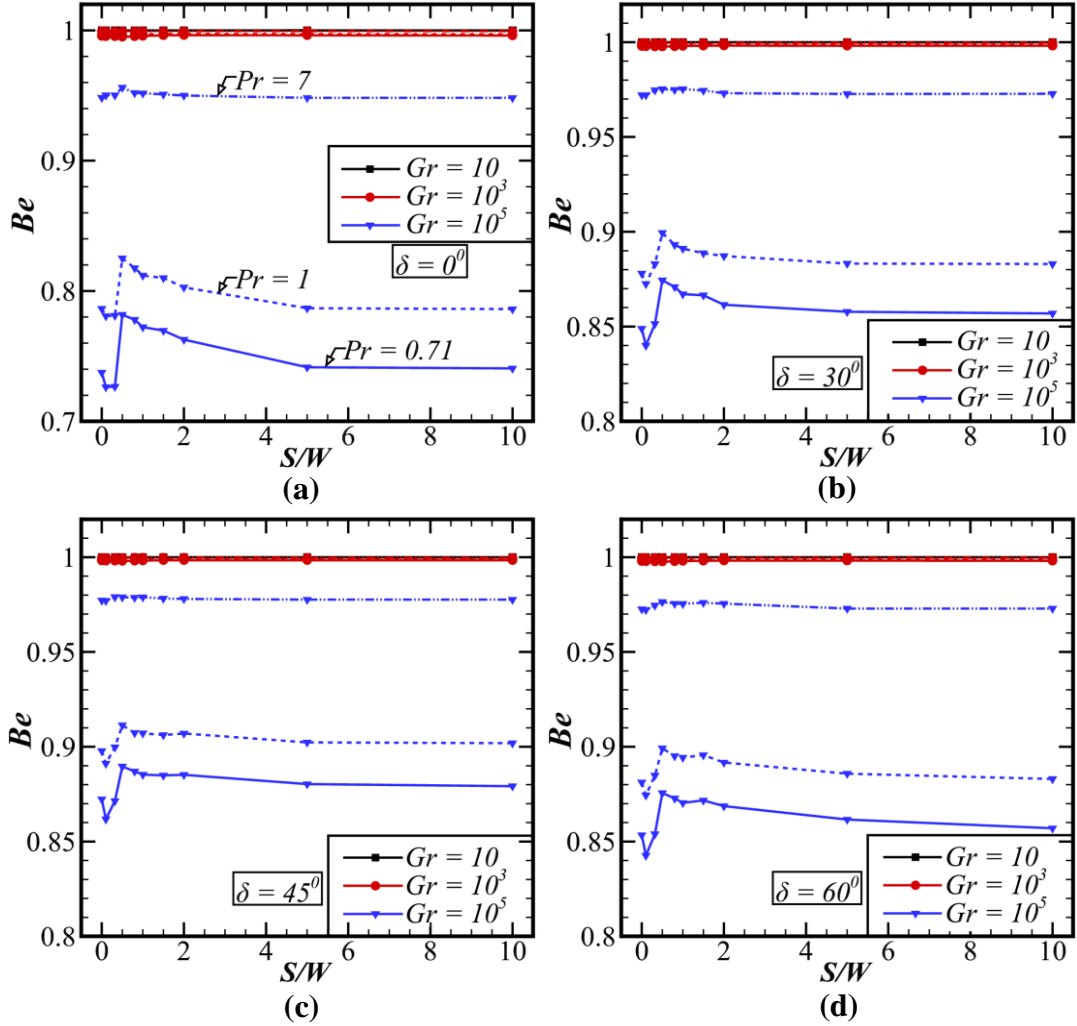

**Figure 18.** Bejan number (*Be*) as a function of *S/W*, *Gr*, and *Pr* (solid line: *Pr* =0.71, dashed line: *Pr* =1, dash-dot line: *Pr* =7) for (a) $\delta=0^0$, (b) $\delta=30^0$, (c) $\delta=45^0$, and (d) $\delta=60^0$

## *4.8. Drag coefficient*

The total drag coefficient, which is the sum of pressure drag and skin-friction drag coefficients is represented in this article to explain the net drag force on the surface on the cylinders. Figure 19 shows the variation of total drag coefficient with horizontal spacing (*S/D*) as a function of *Gr*, *Pr*, and *δ*. It can be seen that due to the formation of a chimney effect, $C_d$ increases with decreases *S/D* up to $S/D_{opt}$, whereas it decreases significantly with further decrease in *S/W*. Owing to the non-dimensional scheme, $C_d$ decreases with increase in *Gr* and *Pr*, albeit the drag force should increase with increase in momentum of the flow. The total drag coefficient ($C_d$) is lower for the square cylinders having tilt angle $\delta=45^0$ compared to $\delta=60^0$ followed by $\delta=30^0$ and $\delta=0^0$ irrespective of $Gr$, $Pr$, and $S/W$. The differences in the value of $C_d$ between various tilt angles are higher at the low value of $Gr$ and $Pr$ as shown in Figure 20. $C_d$ at $\delta=0^0$ is in excess of 30% than the value at $\delta=45^0$.

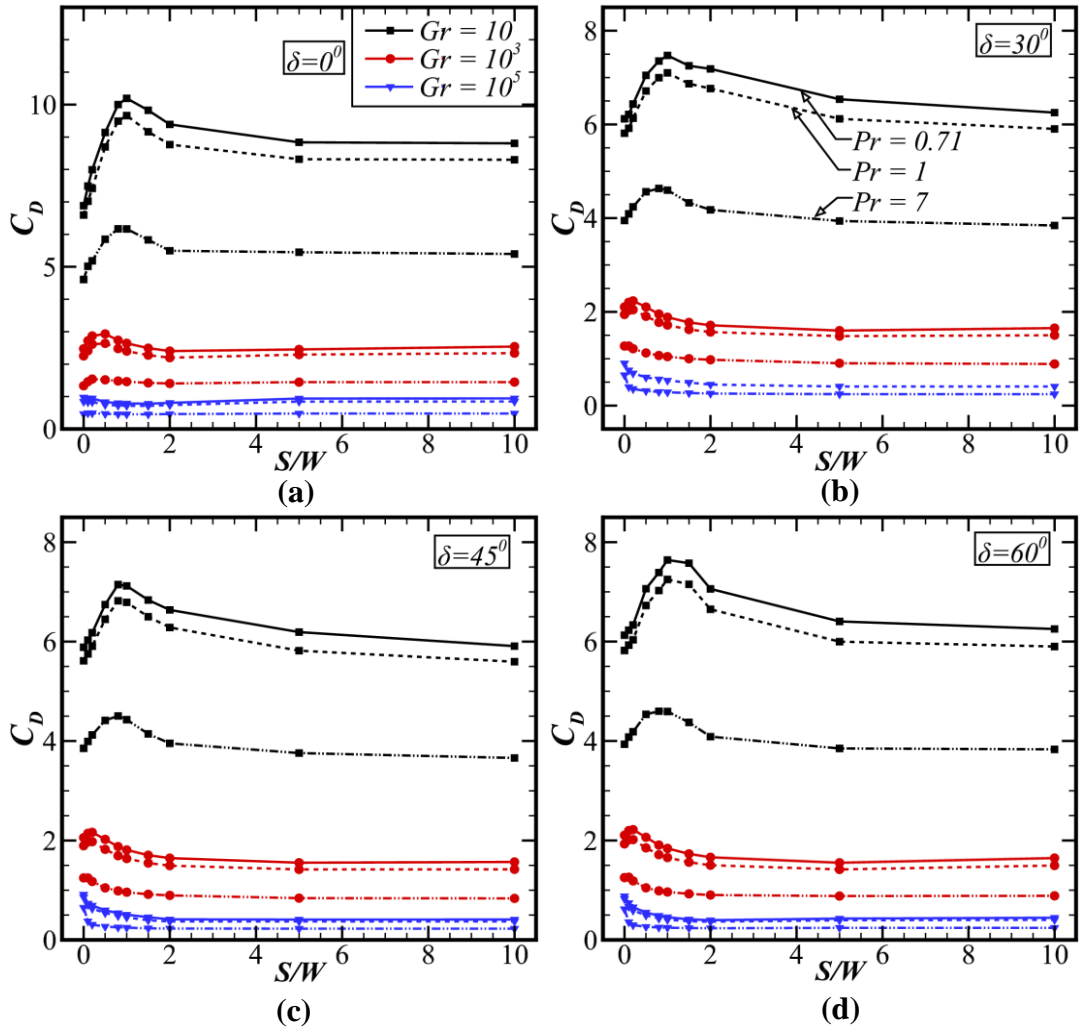

**Figure 19.** Total drag coefficient ($C_D$) as a function of $S/W$, $Gr$, and $Pr$ (solid line: $Pr =0.71$, dashed line: $Pr =1$, dash-dot line: $Pr =7$) for (a) $\delta=0^0$, (b) $\delta=30^0$, (c) $\delta=45^0$, and (d) $\delta=60^0$

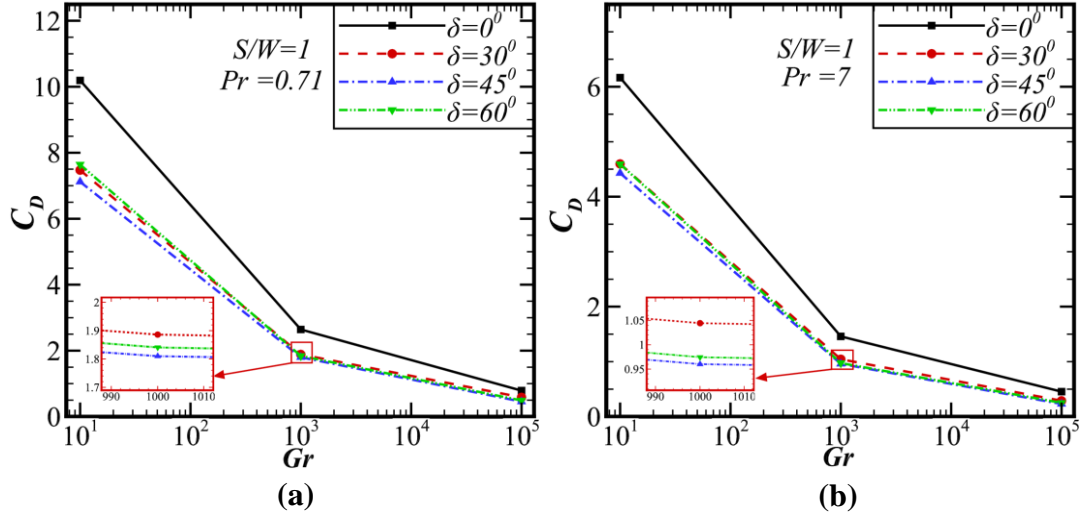

**Figure 20.** Variation of total drag coefficient ($C_D$) with $Gr$ for different tilt angles ($\delta$) of the square cylinder at $S/W=1$ for (a) $Pr=0.71$ and (b) $Pr=7$

## 5. Correlation

An empirical correlation for the average $Nu$ is proposed in the present study by performing a non-linear regression analysis of the numerical data using the Levenberg-Marquardt (LM) algorithm. It has great importance in the engineering design calculations and can be useful to academic researchers and practicing engineers. The average $Nu$ has a functional relationship with $Gr$, $Pr$, $S/W$, and $\delta$. Hence, the correlating equation as expressed in Eq. (32) is valid in the following ranges of independent variables: $10 \leq Gr \leq 10^5$, $0.71 \leq Pr \leq 7$, $0 \leq S/W \leq 10$, and $0^0 \leq \delta \leq 60^0$. The correlation constants and the coefficient of the correlation ($R^2$) of the correlating equation are shown in Table 6. The values of $\delta$ in Eq. (32) are considered to be in radian and the values are given in Table 7. Figure 21 shows a good agreement between the predicted and computed values of average Nu within an accuracy limit of ±8%.

$$Nu = a(Gr \times Pr)^b + c\left[1+(S/W)\right]^d - e(\delta) \quad (32)$$

**Table 6.**
Correlation constants and coefficient of correlation ($R^2$) for the correlating equation

| a | b | c | d | e | $R^2$ |
|---|---|---|---|---|---|
| 0.20643 | 0.30637 | 1.16019 | 0.24036 | 0.66844 | 0.9862 |

**Table 7.**
Values of tilt angle $\delta$ in radian

| $\delta$ | Value (radian) |
|---|---|
| $0^0$ or $90^0$ | 1.5708 |
| $30^0$ | 0.5236 |
| $45^0$ | 0.7854 |
| $60^0$ | 1.0472 |

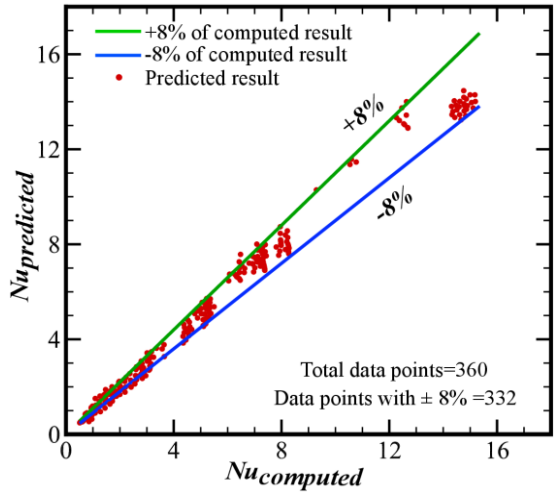

**Figure 21.** Comparison of predicted values of *Nu* from the correlation with the numerically computed values

## 6. Concluding Remarks

In this work, laminar natural convection from two horizontally aligned square cylinders has been investigated numerically to elucidate the effect of tilt angle and horizontal spacing on both momentum and heat transfer characteristics. Computations were performed for the following ranges of pertinent parameters: horizontal spacing: $0 \leq S/W \leq 10$, tilt angle: $0^0 \leq \delta \leq 60^0$, Grashof number: $10 \leq Gr \leq 10^5$, and Prandtl number: $0.71 \leq Pr \leq 7$. The effect on the thermodynamic aspect was also examined in terms of entropy generation in the system. Some of the major findings from this present study are enumerated below.

1. At very low horizontal spacing (*S/W*), a narrow passage or chimney effect is formed, which accelerates the momentum of the thermo-buoyant flow.
2. The thermal boundary layer is gradually becoming thinner with increase in *Gr* and/or *Pr*. Hence, a higher heat transfer is expected at high *Gr* and *Pr*.
3. The velocity and temperature profiles are seen to be steeper corresponding to high *Gr* and/or *Pr*, which indicates a thinner boundary layer over the cylinders.
4. Owing to the formation of a chimney effect, the mass flow rate is higher at the optimum horizontal spacing ($S/W_{opt}$).
5. The average *Nu* is found to show a positive dependence on both *Gr* and *Pr*. Furthermore, it is maximum at the optimal spacing ($S/W_{opt}$) and decreases with increase or decrease in *S/W*.

6. Heat transfer from the square cylinders having tilt angle $\delta=45^0$ is higher than $\delta=60^0$ followed by $\delta=30^0$ and $\delta=0^0$, respectively. At $Gr=10^5$ and $Pr=7$, the average $Nu$ is found to be in excess of 22% at $\delta=45^0$ compared to at $\delta=0^0$.

7. The total drag coefficient ($C_d$) is seen to be maximum at the optimum horizontal spacing and decreases significantly with decrease in *S/W*.

8. Overall, all else being equal, the $C_d$ shows a negative dependence on both *Gr* and *Pr*. It is found to be lower for $\delta=45^0$ and in excess of 30% at $\delta=0^0$.

9. The entropy generation due to fluid flow is seen to be insignificant at low *Gr* and *Pr*. Hence, the Bejan number (*Br*) is almost 100% at low Gr and decreases with increase in *Gr* and *Pr*.

10. An empirical correlation for the average *Nu* is proposed as a function of *Gr*, *Pr*, *S/W*, and *δ*, which would be beneficial to academic researchers and practicing engineers in process industries.

**Acknowledgment**

The present research work was carried out in Computational Fluid Dynamics (CFD) laboratory of the Department of Mechanical Engineering at the Indian Institute of Technology Kharagpur, India.

**Nomenclature**

| | |
|---|---|
| $Be$ | Bejan number |
| $C_d$ | Total drag coefficient |
| $C_p$ | Specific heat at constant pressure, $J/kg\text{-}K$ |
| $g$ | acceleration due to gravity, $m/s^2$ |
| $Gr$ | Grashof number |
| $h$ | heat transfer coefficient, $W/m^2\text{-}K$ |
| $I$ | rate of thermodynamic irreversibility, $W$ |
| $K$ | thermal conductivity, $W/m\text{-}K$ |
| $L$ | characteristic length scale, $m$ |
| $\dot{m}$ | mass flow rate, $kg/s$ |
| $M$ | dimensionless mass flow rate |
| $Nu$ | Nusselt number |
| $p$ | pressure, $N/m^2$ |
| $Pr$ | Prandtl number |
| $S$ | horizontal spacing ($m$) |
| $\dot{s}_{gen}$ | rate of entropy generation per unit volume, $W/m^3\text{-}K$ |
| $T$ | temperature, $K$ |
| $u$ | velocity in $x$-direction, $m/s$ |
| $U$ | velocity magnitude, $m/s$ |
| $v$ | velocity in $y$-direction, $m/s$ |
| $W$ | side of the square cylinder ($m$) |
| $x$ | cartesian $x$-direction |
| $y$ | cartesian $y$-direction |

**Greek Symbols**

| | |
|---|---|
| $\delta$ | tilt angle of the square cylinders, *degree* |
| $\rho$ | density of the fluid, $kg/m^3$ |
| $\beta$ | thermal expansion coefficient, $1/K$ |
| $\mu$ | dynamic viscosity, $kg/m\text{-}s$ |
| $\Phi$ | viscous dissipation per unit volume, $1/s^2$ |
| $\varphi$ | normalized temperature, $K$ |
| $\upsilon$ | kinematic viscosity, $m^2/s$ |
| $\alpha$ | thermal diffusivity, $m^2/s$ |
| $\theta$ | radial angle, *degree* |
| $\psi$ | stream function, $kg/s$ |
| $|\psi|$ | dimensionless stream function; $|\psi| = \psi/\mu L$ |

**Subscripts**

| | |
|---|---|
| $w$ | wall |
| $m$ | mean |
| $\infty$ | ambient |
| $l$ | local |
| $HT$ | heat transfer |
| $FF$ | fluid friction |
| $pr$ | pressure |
| $sf$ | skin-friction |
| $opt$ | optimum |